\documentclass[a4paper,11pt]{article}

\usepackage{jheppub}
\usepackage{xcolor}

\usepackage{mathrsfs}
\usepackage{amsmath,amsthm,amssymb,slashed}
\usepackage{color}
\usepackage{multirow}
\usepackage[normalem]{ulem}
\usepackage{afterpage}

\newcommand{\cm}[1]{{\color{black}#1}}
\makeatletter
\providecommand{\@fpheader}{}
\makeatother

\title{\cm{Investigating DUNE oscillations sensitivity to sterile Pseudo-Dirac Neutrinos}
}
\author[a]{Asmaa Abada,}
\affiliation[a]{Pôle Théorie, Laboratoire de Physique des 2 Infinis Irène Joliot-Curie (UMR 9012), \\
  CNRS/IN2P3, 15 Rue Georges Clemenceau, 91400 Orsay, France}
\emailAdd{asmaa.abada@ijclab.in2p3.fr}

\author[b]{Jo\~ao Paulo Pinheiro,}
\affiliation[b]{Departament de Física Quàntica i Astrofísica and Institut de Ciències del Cosmos, \\
  Universitat de Barcelona, Diagonal 647, E-08028 Barcelona, Spain}
\emailAdd{joaopaulo.pinheiro@fqa.ub.edu}

\author[a]{Salvador Urrea}
\emailAdd{salvador.urrea@ijclab.in2p3.fr}

\abstract{We explore the sensitivity of the Deep Underground Neutrino Experiment (DUNE) to sterile neutrino oscillations within a $3+$(pseudo-Dirac pair) framework. We first consider a pair of two sterile neutrinos forming a pseudo-Dirac pair, then we consider a low-scale seesaw realization, that we name ``Linear-Inverse Seesaw" 
model. This scenario features two nearly degenerate sterile neutrino states at the keV scale, characterized by a small mass splitting arising from a small amount of lepton number violation.

In this scenario, the oscillation behavior can be described in three distinct regimes depending on the sterile-sterile mass-squared difference: low ($< 1\,\mathrm{eV}^2$), resonant ($1$--$100\,\mathrm{eV}^2$), and high ($> 100\,\mathrm{eV}^2$) regimes, recovering in both low- and high-mass regimes an effective non-unitarity of the %active 
leptonic mixing matrix. A distinctive feature of this framework is that observable effects persist even in the low-mass limit, unlike the case of standard $3+1$ scenarios, due to rapid oscillation averaging from larger keV-scale splittings.

We leverage the complementarity of both near  
and far detectors to explore the sensitivity to $\nu_e$ and $\nu_\mu$ disappearance and $\nu_e$ and $\nu_{\tau}$ appearance oscillation probabilities. Our analysis reveals that DUNE can achieve significant improvements over current experimental constraints, especially in neutrino appearance modes. Additionally, we show that new CP-violating phases associated with the sterile sector can dramatically alter the sensitivity, with destructive interference potentially suppressing signals by orders of magnitude.}


\begin{document}

\maketitle

\section{Introduction}
\label{sec:intro}

The observation of neutrino oscillations provides clear evidence for lepton flavor violation in the neutral sector and establishes that neutrinos are massive particles and mix. This phenomenon, unexplained within the Standard Model (SM) of particle physics, points unequivocally toward physics beyond the SM (BSM). Although the conventional three-neutrino oscillation framework consistently explains most experimental results~\cite{Esteban:2024eli}, persistent anomalies in short-baseline neutrino experiments have sparked significant interest in the possible existence of additional neutrino states, sterile neutrinos, that do not couple via the standard electroweak interaction.

The strongest experimental indications for the existence of light sterile neutrinos originate from the LSND~\cite{LSND:1995lje} and MiniBooNE~\cite{MiniBooNE:2020pnu} experiments, both of which reported excesses of electron neutrino-like events that cannot be accommodated by the standard three-neutrino scenario. \cm{Furthermore, the so-called Gallium anomaly~\cite{Kaether:2010ag,Abdurashitov:2005tb,GALLEX:1997lja,SAGE:2009eeu,SAGE:1998fvr}, characterized by deficits in measured neutrino fluxes compared to theoretical expectations, remains a long-standing open puzzle. Recent results from the Baksan Experiment on Sterile Transitions (BEST)~\cite{Barinov:2021asz} have confirmed the anomaly at above $5\sigma$, strengthening the case for eV-scale sterile neutrinos. These anomalous observations imply relatively large mixing with the active neutrinos, a scenario strongly constrained by solar neutrino data~\cite{Gonzalez-Garcia:2024hmf}.}

From a theoretical standpoint, sterile neutrinos arise naturally in numerous neutrino mass generation mechanisms aiming to explain the observed smallness of active neutrino masses and the leptonic mixings. The minimal extension of the SM involves the addition of right-handed (RH) neutrinos, leading either to Dirac or Majorana physical (mass) states. 

The canonical Type-I seesaw mechanism~\cite{Minkowski:1977sc,Yanagida:1979as,Mohapatra:1979ia,GellMann:1980vs} incorporates extremely heavy Majorana neutrinos to generate tiny active neutrino masses. Due to the heaviness of the sterile state, it is a challenge to search for direct phenomenological signatures of the heavy neutrinos, as their masses typically lie far beyond the reach of current (and future) experiments. 
 
Several low-energy scale variants of the latter mechanism have been proposed allowing for experimentally accessible heavy neutrino states, while keeping natural small masses for the active ones. The linear seesaw (LSS)~\cite{Akhmedov:1995ip, Barr:2003nn, Malinsky:2005bi} modifies the mass matrix structure of the standard seesaw,  allowing TeV sterile masses or below, and the active neutrino masses proportional to a small symmetry-breaking scale and linear dependency on the Dirac mass. Similarly, the inverse seesaw (ISS) \cite{Mohapatra:1986bd,Gonzalez-Garcia:1988okv,Gavela:2009cd,BERNABEU1987303,Mohapatra:1986aw} introduces additional sterile fermion states and relies on a small lepton number violation energy scale to generate small masses for the active states and TeV sterile neutrinos. In order to comply with neutrino oscillation data, both  ISS and LSS mechanisms necessitate the introduction of at least four sterile neutrino states, both leading to two pseudo-Dirac pairs in the neutrino heavy spectrum.   When these two mechanisms are combined, an especially attractive scenario emerges, the Linear-Inverse Seesaw (LISS) model~\cite{Abada:2015rta}.  In this framework, which requires at least the extension of the SM by two neutral leptons, the heavy spectrum presents only one pseudo-Dirac pair,  
whose mass can range from 1~keV to a few TeV and its  mass splitting controlled by a small lepton number violation, can accommodate the $\sim 1\mathrm{eV}^2$ mass differences typically probed in short-baseline experiments or near detectors of long-baseline facilities. \cm{To avoid confusion with other scenarios in the literature involving pseudo-Dirac pairs, it is important to clarify at this point that we consider the effect of pseudo-Dirac sterile neutrinos on neutrino oscillation probabilities. We do not assume the active neutrinos to be of pseudo-Dirac nature~\cite{Wolfenstein:1981rk,Petcov:1982ya,Valle:1983dk}, nor do we study the pseudo-Dirac scenario for active neutrino oscillations~\cite{Kobayashi:2000md}. In our model, the active neutrinos are Majorana particles due to the small violation of lepton number.}

Beyond their role in neutrino oscillations, sterile neutrinos have significant implications across cosmology and astrophysics, potentially addressing questions such as dark matter nature and baryon asymmetry via leptogenesis~\cite{Abazajian:2012pn,Abazajian:2017tcc}. \cm{In particular, keV-scale sterile neutrinos could play an important role in supernova evolution, particularly during the cooling phase, as discussed in~\cite{Suliga:2019bsq,Suliga:2020vpz}. }However, sterile neutrinos below the GeV scale face stringent cosmological constraints: masses under approximately 100 MeV are strongly limited by structure formation and Lyman-$\alpha$ forest data~\cite{Smirnov:2006bu,Kusenko:2009up,Drewes:2016upu}, while radiative decays ($\nu_i \rightarrow \nu_j \gamma$) are severely restricted by cosmic X-ray observations~\cite{Loewenstein:2008yi,Loewenstein:2012px}. At even lower masses, constraints from Big Bang Nucleosynthesis (BBN) and Cosmic Microwave Background (CMB) measurements become crucial, as sterile neutrinos decaying after the onset of BBN can modify the primordial helium abundance and the effective number of neutrino species ($N_{\text{eff}}$)~\cite{Kawasaki:2000en}. Dedicated studies have systematically explored these constraints, demonstrating their severity below the $\mathcal{O}(100)$~MeV scale~\cite{Hernandez:2014fha}. Nevertheless, these cosmological bounds can be circumvented through non-standard cosmological scenarios~\cite{Gelmini:2004ah,Gelmini:2008fq}, hidden-sector couplings~\cite{Bezrukov:2017ike,Dasgupta:2013zpn}, extended gauge frameworks such as Left-Right symmetric models~\cite{Barry:2014ika}. Another alternative arises in models with an extended Higgs sector, where the Majorana mass term of sterile neutrinos can originate from the vacuum expectation value of a gauge-singlet scalar field~\cite{Petraki:2007gq}. Such scenarios decouple sterile neutrino relic abundances from active-sterile mixing angles, reducing the free-streaming length compared to standard neutrino oscillation production mechanisms (e.g., the Dodelson-Widrow mechanism~\cite{Dodelson:1993je}), and thus significantly relaxing cosmological and X-ray constraints.

At higher sterile neutrino mass scales (above tens of MeV and up to $\sim$ 100 GeV), direct laboratory searches become particularly relevant. Extensive experimental efforts have searched for signatures of sterile neutrino production and decay across various experiments, placing significant constraints on their mixing with active neutrinos~\cite{Shrock:1980vy, Shrock:1981wq, Atre:2009rg,Abada:2013aba, Abada:2017jjx,Deppisch:2015qwa,Alekhin:2015byh,Antel:2023hkf}. Specifically, direct detection experiments sensitive to Majorana sterile neutrino signatures in meson decays and high-energy colliders have systematically probed the parameter space, ruling out significant portions of active-sterile mixing for neutrino masses in the MeV–GeV range~\cite{Abada:2017jjx}.

In the intermediate mass regime around the keV scale, sterile neutrinos could manifest themselves through distortions in the electron energy spectrum of tritium beta-decay experiments, providing a complementary and cosmologically independent probe of their existence~\cite{Shrock:1980vy}. The ongoing KATRIN experiment, along with its forthcoming TRISTAN upgrade~\cite{KATRIN:2001ttj,Mertens:2014nha,Boyarsky:2018tvu,Abada:2018qok}, is specifically designed to detect such spectral deviations. To date, these laboratory searches have not observed sterile neutrino signals, placing stringent constraints on active-electron neutrino mixing angles at keV-scale masses.

For sterile neutrinos with masses below the sensitivity threshold of beta-decay experiments, or in scenarios that evade cosmological and astrophysical constraints as described earlier, neutrino oscillation experiments represent the primary available detection pathway. \cm{In this work, we focus on the experimental sensitivity of future long-baseline experiments to new physics. Among the next-generation facilities, the Deep Underground Neutrino Experiment (DUNE)~\cite{DUNE:2020ypp,DUNE:2020jqi} and Tokai to Hyper-Kamiokande (T2HK)~\cite{Abe:2015zbg,Abe:2018uyc} are expected to play leading roles in the exploration of neutrino properties. For sterile neutrino searches in particular, the sensitivity is largely driven by the performance of the near detector. Among the near detectors planned for T2HK, the only one with a fiducial mass comparable to DUNE’s near detector is the intermediate water-Cherenkov detector at about 1~km from the target~\cite{nuPRISM:2014mzw,Hyper-Kamiokande:2018ofw,HyperKamiokande2025ESPP}. While this detector will indeed be larger than the DUNE near detector, the dominant sensitivity to sterile neutrinos arises from spectral energy information. In this respect, the T2HK beam, produced by 30~GeV protons and operated off-axis~\cite{T2K:2011qtm}, has substantially lower energy than the 120~GeV proton beam at DUNE~\cite{DUNE:2020lwj}. 
%Moreover, water-Cherenkov detectors suffer from poorer energy resolution and less effective track reconstruction compared to liquid argon time-projection chambers, both of which are crucial to disentangle potential new-physics effects from the Standard Model background. 
For these reasons, we expect DUNE to provide stronger sensitivity than T2HK in sterile neutrino searches.
}

%Among the next-generation long-baseline experiments, the Deep Underground Neutrino Experiment (DUNE)\cite{DUNE:2020ypp,DUNE:2020jqi} and Tokai to Hyper-Kamiokande (T2HK)\cite{Abe:2015zbg,Abe:2018uyc} are poised to play leading roles in the exploration of neutrino properties and searches for new physics. Featuring a high-intensity neutrino beam and detectors placed at two baselines (a Near Detector at 574~m and a Far Detector at 1300~km), DUNE is uniquely positioned to precisely measure both appearance and disappearance channels. Consequently, it will offer unprecedented sensitivity to sterile neutrino oscillation signatures.

Motivated by the above considerations, this study investigates DUNE’s sensitivity to sterile neutrino oscillations in a 3+(pseudo-Dirac pair) scenario, specifically within the context of the Linear Inverse Seesaw model. We extend the standard three-neutrino oscillation formalism by introducing two additional sterile neutrino states that form a quasi-degenerate pair. These states are characterized by two distinct regimes for their mass splitting: a relatively large splitting of order $\mathrm{keV}^2$, which roughly matches the squared masses of the heavy species, small enough to evade direct-detection limits yet large enough to relax cosmological bounds, and a tiny splitting of order $\mathrm{eV}^2$ between the sterile states themselves, generated by small lepton-number violation. Such a hierarchical mass structure emerges naturally within the LISS framework, resulting in distinctive oscillation signatures that can be robustly probed by DUNE’s experimental capabilities.

The paper is structured as follows: Section~\ref{sec:oscillation_LISS} describes the oscillation framework for the 3+(pseudo-Dirac pair) scenario, deriving oscillation probabilities relevant to both near and far detector setups. In Section~\ref{sec:LISS}, we introduce the theoretical details of the LISS model and explore its parameter space consistent with current neutrino oscillation constraints. Section~\ref{sec:exp_setup} outlines the simulation methods and the experimental setup used to evaluate DUNE’s sensitivity. Results are presented and analyzed in Section~\ref{sec:results}, highlighting sensitivity projections, complementarity between near and far detectors, and the role of new CP-violating phases. Finally, conclusions and perspectives are summarized in Section~\ref{sec:conclusions}.

\section{Oscillations in a 3+(pseudo-Dirac pair) scenario}
\label{sec:oscillation_LISS}

In this section, we explore neutrino oscillations within a minimal extension of the standard three-neutrino framework. Specifically, we consider the addition of two sterile neutrino states forming a nearly degenerate pseudo-Dirac pair, a scenario naturally emerging in low-scale neutrino mass generation mechanisms, notably the Linear Inverse Seesaw (LISS, discussed in detail in Section~\ref{sec:LISS}). This extension leads to a 3+(pseudo-Dirac pair) framework, where the leptonic mixing matrix is enlarged from the standard $3\times3$ matrix to a $5\times5$ matrix:

\begin{equation}
U = \begin{pmatrix}
U_{e1} & U_{e2} & U_{e3} & U_{e4} & U_{e5} \\
U_{\mu 1} & U_{\mu 2} & U_{\mu 3} & U_{\mu 4} & U_{\mu 5} \\
U_{\tau 1} & U_{\tau 2} & U_{\tau 3} & U_{\tau 4} & U_{\tau 5} \\
U_{s1,1} & U_{s1,2} & U_{s1,3} & U_{s1,4} & U_{s1,5} \\
U_{s2,1} & U_{s2,2} & U_{s2,3} & U_{s2,4} & U_{s2,5}
\end{pmatrix}.
\label{eq:55matrix}
\end{equation}

The upper-left $3\times3$ submatrix ($U_{ei}, U_{\mu i}, U_{\tau i}$ with $i=1,2,3$) corresponds approximately to the standard leptonic mixing matrix ($U_\text{PMNS}$)%($U_\text{standard}$)
, whereas the matrix elements $U_{ek}, U_{\mu k}, U_{\tau k}$ (for $k=4,5$) parametrize the mixing between active and sterile neutrino states. The leptonic mixing matrix is commonly parametrized in terms of mixing angles ($\theta_{ij}$), Dirac and Majorana phases ($\delta_{ij}$ and $\varphi_i$, respectively), as detailed in Appendix~\ref{App:Parametrization_leptonic_mixing}.
The last two lines of the matrix cannot be directly probed. 

Under constant matter density, neutrino propagation is governed by the effective Hamiltonian in the flavor basis:
\cm{
\begin{equation}\label{eq:H_mass_basis}
H = \frac{1}{2E}\,U\mathrm{diag}(m_1^2, m_2^2, m_3^2, m_4^2, m_5^2)U^{\dagger}
+ \mathrm{diag}(V_\text{NC}+V_\text{CC},V_\text{NC},V_\text{NC},0,0)\,,
\end{equation}}
\cm{where $E$ is the neutrino energy, $m_i$ are the physical neutrino masses ($i=1,\dots,5$), $V_{\text{CC}}$
 is the charged-current matter potential for electron (anti)neutrinos, $V_{\text{CC}}=\pm\sqrt{2}\,G_F N_e$, and 
$V_{\text{NC}}=-\frac{G_F}{\sqrt{2}}\,N_n$ is the neutral-current matter potential common to all active (anti)neutrinos. 
Here, $G_F$ is the Fermi constant, $N_e$ the electron number density, and $N_n$ the neutron number density. 
The first term in Eq.~\eqref{eq:H_mass_basis} is the vacuum neutrino Hamiltonian, while the second term encodes coherent forward scattering in matter: charged-current scattering of $\nu_e$ on electrons and 
neutral-current scattering of the active neutrinos on neutrons.}
%For the Earth's crust, the neutron-to-electron density ratio is approximately constant, $Y \equiv N_n/N_e \simeq 1.05$~\cite{ADDREF}. 

The neutrino evolution operator is formally expressed as:
\begin{equation}
S(t)=\exp(-iHt)\,.
\end{equation}
Diagonalizing the Hamiltonian as $H=\tilde{U}\Lambda\tilde{U}^{\dagger}$ (with $\Lambda=\operatorname{diag}\left(\lambda_1,\lambda_2,\lambda_3,\lambda_4,\lambda_5\right)$)\footnote{We present the derivation explicitly in the flavor basis for clarity, though our numerical implementation employs the mass basis.}, allows the evolution operator to be expressed as:

\begin{equation}
S(t)=\tilde{U}\exp(-i\Lambda t)\tilde{U}^{\dagger}\,.
\end{equation}

Thus, the transition probability from an initial flavor $\nu_{\beta}$ to a final flavor $\nu_{\alpha}$ is given by:
\begin{equation}\label{eq:transition_matter_effects}
P_{{\nu}_{\beta} \rightarrow \nu_{\alpha}}(t)=\left|\langle\nu_\alpha|S(t)|\nu_\beta\rangle\right|^2=\left|\sum_{k=1}^5 \tilde{U}_{\alpha k}\tilde{U}_{\beta k}^* e^{-i\lambda_k t}\right|^2\,.
\end{equation}

Expanding, we find:

\begin{equation}\label{eq:probability_in_matter}
\begin{aligned}
P_{\nu_{\beta}\rightarrow \nu_{\alpha}}(t)= &\sum_{k=1}^5|\tilde{U}_{\alpha k}|^2|\tilde{U}_{\beta k}|^2\\
&+2\sum_{k>j}\Big\{\Re\left[\tilde{U}_{\alpha k}\tilde{U}_{\beta k}^*\tilde{U}_{\alpha j}^*\tilde{U}_{\beta j}\right]\cos\left[(\lambda_k-\lambda_j)t\right]\\
&\quad -\Im\left[\tilde{U}_{\alpha k}\tilde{U}_{\beta k}^*\tilde{U}_{\alpha j}^*\tilde{U}_{\beta j}\right]\sin\left[(\lambda_k-\lambda_j)t\right]\Big\}\,.
\end{aligned}
\end{equation}

Although analytically exact, Eq.~\eqref{eq:probability_in_matter} is not directly observable experimentally. In realistic scenarios, finite detector energy resolution necessitates averaging the transition probability over an appropriate energy distribution. Applying a Gaussian filter that models finite detector resolution~\cite{Huber:2004ka,Huber:2007ji,Coloma:2021uhq}, we obtain the experimentally observable probability:

\begin{equation}
\begin{aligned}
\langle P_{\nu_{\beta}\rightarrow\nu_{\alpha}}(t)\rangle=&\sum_{k=1}^5|\tilde{U}_{\alpha k}|^2|\tilde{U}_{\beta k}|^2\\
&+2\sum_{k>j}f_{kj}\Big\{\Re\left[\tilde{U}_{\alpha k}\tilde{U}_{\beta k}^*\tilde{U}_{\alpha j}^*\tilde{U}_{\beta j}\right]\cos\left[(\lambda_k-\lambda_j)t\right]\\
&\quad-\Im\left[\tilde{U}_{\alpha k}\tilde{U}_{\beta k}^*\tilde{U}_{\alpha j}^*\tilde{U}_{\beta j}\right]\sin\left[(\lambda_k-\lambda_j)t\right]\Big\}\,,
\end{aligned}
\end{equation}
where the Gaussian damping factor is $f_{kj}=\exp\left[-\frac{1}{2}\left(\frac{\sigma_E}{E}\right)^2\left((\lambda_k-\lambda_j)t\right)^2\right]$, and $\sigma_E$ denotes the detector energy resolution.

In this work, we assume two nearly degenerate sterile neutrinos with masses $\sim$ keV, %that are nearly degenerate, 
such that $\Delta m^2_{41}\approx\Delta m^2_{51}\approx 1\,\text{keV}^2$ with a relatively small mass splitting $\Delta m^2_{54}\approx 1-10\,\text{eV}^2$. The small mass-squared difference produces oscillation patterns matching the energy-to-baseline ratio ($E/L$) accessible for the DUNE ND, enabling potential observation of these oscillations. Conversely, the larger mass-squared differences $\Delta m_{41}^2,\, \Delta m_{51}^2$ will average out at the ND but contribute as constant terms to the transition probabilities. At longer baselines (DUNE FD), both the small and large mass-squared differences will have undergone many oscillation cycles. While the large differences remain averaged out, the small splitting $\Delta m^2_{54}$ may also begin to average when $\Delta m^2_{54}>0.003$ eV$^2$, for the typical energies of the DUNE experiment (where $90\%$ of the incident flux of $\nu_\mu$ have energies between 0.5 and 6 GeV).

\subsection{Near Detector}
\label{subsec:ND_theory}
For the ND ($L\approx 500$ m), there is not sufficient baseline for the active neutrinos to undergo standard oscillation.  At the same time, for this distance, matter effects become negligible, and the oscillation probabilities simplify considerably. In this limit, Eq.~\eqref{eq:transition_matter_effects} reduces to:

\begin{equation}
P_{\nu_{\beta}\rightarrow\nu_{\alpha}}=\left|\sum_jU_{\alpha j}U_{\beta j}^*e^{-\frac{i\Delta m_{j1}^2L}{2E}}\right|^2.
\end{equation}

Standard oscillation frequencies ($\Delta m_{21}^2,\Delta m_{31}^2$) effectively vanish. However, mixing between active and sterile neutrinos can lead to the disappearance or appearance of active neutrino states at short baselines. The disappearance probability is expressed as:

\begin{eqnarray}
P_{\nu_{\alpha}\rightarrow\nu_{\alpha}}&=&1-4|U_{\alpha4}|^2(1-|U_{\alpha4}|^2)\sin^2\frac{\Delta_{41}}{2}-4|U_{\alpha5}|^2(1-|U_{\alpha5}|^2)\sin^2\frac{\Delta_{51}}{2}\\
&&+4|U_{\alpha4}|^2|U_{\alpha5}|^2\left(\sin^2\frac{\Delta_{41}}{2}+\sin^2\frac{\Delta_{51}}{2}-\sin^2\frac{\Delta_{41}-\Delta_{51}}{2}\right),\nonumber
\label{eq:dis_prob}
\end{eqnarray}
while the appearance probability is given by:
\begin{eqnarray}
P_{\nu_{\alpha}\rightarrow\nu_{\beta}}&=&4|U_{\alpha4}|^2|U_{\beta4}|^2\sin^2\frac{\Delta_{41}}{2}+4|U_{\alpha5}|^2|U_{\beta5}|^2\sin^2\frac{\Delta_{51}}{2}\\
&&+4\mathrm{Re}[U_{\alpha4}U_{\alpha5}U_{\beta4}^*U_{\beta5}^*]\left(\sin^2\frac{\Delta_{41}}{2}+\sin^2\frac{\Delta_{51}}{2}-\sin^2\frac{\Delta_{41}-\Delta_{51}}{2}\right),\nonumber
\label{eq:app_prob}
\end{eqnarray}
where $\Delta_{ij}=\frac{\Delta m^2_{ij}L}{2E}$.

\begin{figure}[h!]
\centering
\begin{tabular}{cc}
\includegraphics[width=0.47\textwidth]{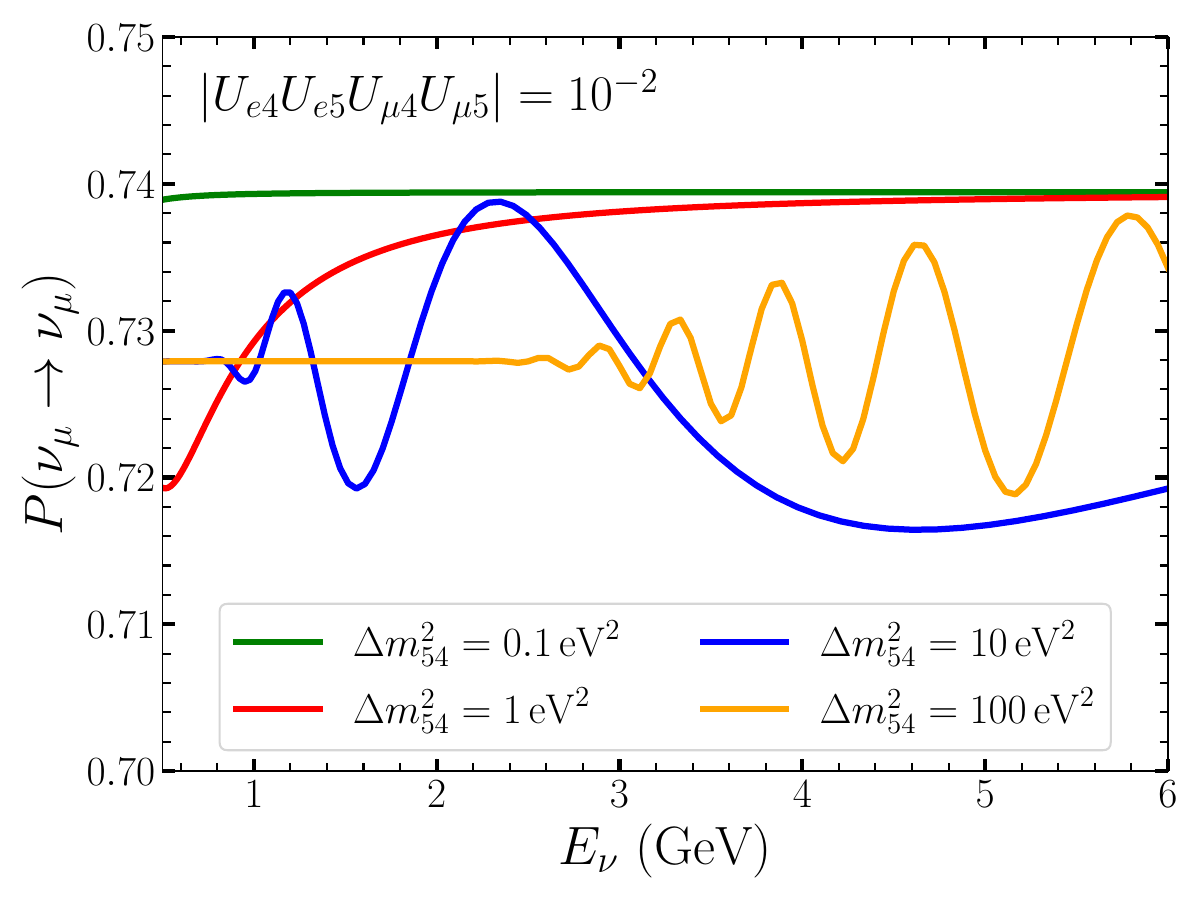}&
\includegraphics[width=0.47\textwidth]{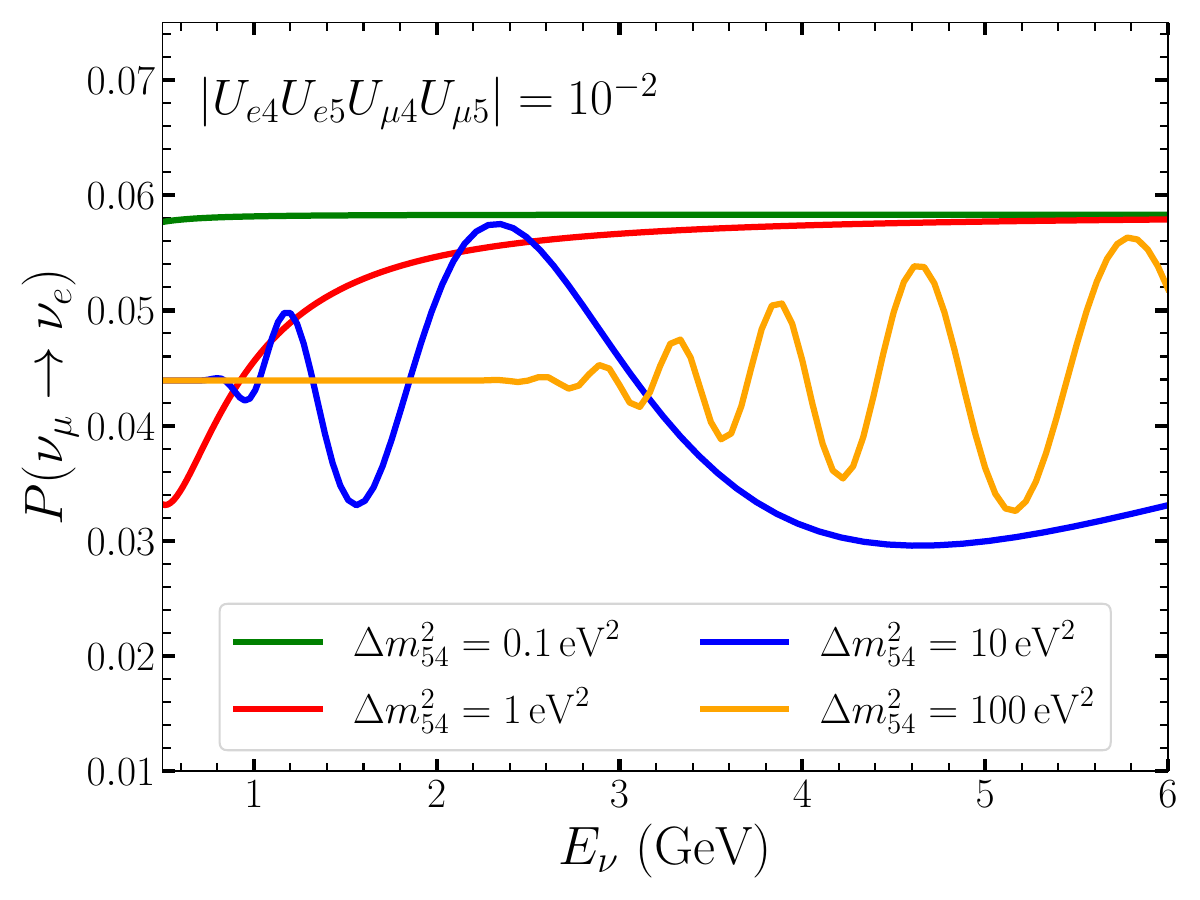}
\end{tabular}
\caption{Muon neutrino disappearance {(left)} and electron neutrino appearance {(right)} probabilities as functions of energy for fixed active-sterile mixing parameters ($|U_{e4}|=|U_{e5}|=|U_{\mu 4}|=|U_{\mu 5}|=0.316$) and varying sterile neutrino mass splittings: $\Delta m_{54}^2 = 0.1$ eV$^2$ (green), $1$ eV$^2$ (red), $10$ eV$^2$ (blue), and $100$ eV$^2$ (orange). The energy range corresponds to more than 90$\%$ of the initial $\nu_\mu$ flux at the DUNE near detector.}
\label{fig:prob_app_dis}
\end{figure}

Figure~\ref{fig:prob_app_dis} illustrates the impact of active-sterile mixing on muon neutrino oscillation probabilities at short baselines, as described by Eqs.~\eqref{eq:dis_prob} and~\eqref{eq:app_prob}. The calculation assumes (for illustration) fixed mixing parameters ($|U_{e4}|=|U_{e5}|=|U_{\mu 4}|=|U_{\mu 5}|=0.316$) across the energy range of 0.5–6.0 GeV.
The oscillation behavior exhibits three distinct regimes as a function of $\Delta m_{54}^2$. In the low-mass regime ($\Delta m_{54}^2 = 0.1$ eV$^2$, green curves), the oscillation phase $\Delta_{54}$ remains small throughout the energy range, causing $\sin^2(\Delta_{54}/2) \to 0$. However, 
the constraints from the ND do not vanish, in contrast to what occurs in the $3+1$ scenario. This behavior originates 
from rapid oscillation terms with phases proportional to $\Delta m_{41}^2$ and $\Delta m_{51}^2$, which average out 
and modify the transition probabilities by a constant offset. Consequently, both appearance and disappearance probabilities approach constant values with minimal energy dependence.

The resonant oscillation regime emerges for intermediate mass-squared differences ($\Delta m_{54}^2 = 1$ and $10$ eV$^2$, red and blue curves respectively). Here, the phase $\Delta_{54}$ leads to $\sin^2(\Delta_{54}/2) \to 1$ and produces significant oscillatory behavior. The neutrino survival and conversion probabilities exhibit clear energy-dependent modulations characteristic of two-neutrino mixing between the nearly degenerate sterile states.

At larger mass-squared differences ($\Delta m_{54}^2 = 100$ eV$^2$, orange curves), the system enters the average out regime. The rapid phase evolution results in high-frequency oscillations that exceed typical detector energy resolution, causing $\sin^2(\Delta_{54}/2)$ to average to 1/2. This averaging effectively washes out the oscillatory signatures.
 
Following the above discussion, the sensitivity to the disappearance/appearance channels can be understood by examining three 
distinct regimes of $\Delta m_{54}^2$:

\begin{itemize}
\item \textbf{Low-mass regime} ($\Delta m_{54}^2 < 1\,\text{eV}^2$): 
      The pseudo-Dirac mass splitting is not yet developed at the ND .
      
\item \textbf{Resonant regime} ($1\,\text{eV}^2 < \Delta m_{54}^2 < 70-100\,\text{eV}^2$): 
      Best sensitivity at the ND where $\sin^2(\Delta_{54}/2) \to 1$, maximizing the energy-dependent features of the oscillation probability.
      
\item \textbf{High-mass regime} ($\Delta m_{54}^2 > 100\,\text{eV}^2$): 
      Rapid oscillations averaging out the pseudo-Dirac mass splitting.
\end{itemize}

An important observation is that both the low-mass and high-mass regimes result in constant deviations from unitarity in the oscillation probabilities. These limiting cases can therefore be interpreted as manifestations of low-scale non-unitarity in the active block of the leptonic mixing matrix (see Refs.~\cite{Blennow:2025qgd,Blennow:2023mqx}).\footnote{This behavior differs from the well-studied 3+1 sterile neutrino scenario, where only the high-mass regime induces non-unitarity effects, while the low-mass regime recovers the SM prediction.} At leading order in disappearance oscillation probability, both regimes coincide, whereas in the appearance channel,  $P^ {\text{High-mass}}_{\nu_{\alpha}\rightarrow\nu_{\beta}} \approx \frac{3}{4} P^ {\text{Low-mass}}_{\nu_{\alpha}\rightarrow\nu_{\beta}} $  in the limit of democratic mixing ($|U_{\alpha 4}|=|U_{\alpha 5}|=|U_{\beta 4}|=|U_{\beta 5}|$).

\section{Linear inverse seesaw (LISS)}
\label{sec:LISS}

One realization of the described scenario in Section~\ref{sec:oscillation_LISS}, featuring two $\sim$keV sterile neutrinos quasi-degenerate in mass, is the Linear Inverse Seesaw (LISS) mechanism. As mentioned in the introduction, the LISS combines elements from both the Inverse and Linear seesaw models, providing a framework that generates the required mass spectrum and mixing patterns.

The LISS most basic realization is by an extension of the SM by introducing two right-handed neutrinos~\cite{Gavela:2009cd,Abada:2015rta}, $N_R^1$ and $N_R^2$, at a new physics scale $\Lambda$, each carrying opposite lepton numbers. Specifically, the lepton number assignments for these fields are $L(N_R^1) = +1$ and $L(N_R^2) = -1$. This mechanism relies on an approximate lepton number symmetry, meaning that the neutrino masses remain naturally small as their size is proportional to small lepton number violating (LNV) parameters. The smallness of these parameters is technically natural in the sense of 't Hooft~\cite{tHooft:1979rat}, as setting them to zero restores an exact lepton number symmetry, protecting neutrino masses from large radiative corrections.

\subsection{The model}
\label{subsec:model_LISS}

The simplest Lagrangian (combined with the SM Lagrangian, ${\cal L}_\text{SM}$) that incorporates these right-handed neutrinos is:
\begin{eqnarray}
    {\cal L} = {\cal L}_\text{SM} + i \bar{N}_R^1 \slashed{\partial}N_R^1+ i \bar{N}_R^2 \slashed{\partial}N_R^2-( Y_{\alpha }\bar{\ell}_\alpha \Tilde{\phi}N_R^1 + \epsilon Y^\prime_{\alpha } \bar{\ell}_\alpha \Tilde{\phi}N_R^2 + \frac{\Lambda}{2}\bar{N}^{1c}_R N_R^2 +\frac{\xi \Lambda}{2}\bar{N}^{2c}_R N_R^2 ),\nonumber\\ 
\end{eqnarray}
such that the dimensionless parameters $\epsilon$ and $\xi$ are supposed to be small and quantify the amount of lepton number violation. The global neutrino mass term, then, reads:
\begin{eqnarray}
    -{\cal L}_{m_\nu}=n_L^T C {\cal M}n_L + h.c.,
\end{eqnarray}
where
$n_L=(\nu_{eL},\nu_{\mu L},\nu_{\tau L}, {N_1}^c, {N_2}^c)^T$ and $C=i \gamma^2\gamma^0$ and the mass matrix is given by
\begin{equation}\label{eq:MLISS}
{\cal M} =  \left( \begin{array}{ccc}
 \mathbf{0} & \textbf{Y} v &\epsilon \textbf{Y}' v\\
 \textbf{Y}^T v & 0 & \Lambda \\
 \epsilon{\textbf{Y}}'^T v & \Lambda & \mu 
 \end{array}
\right),
\end{equation}
where $\mu\equiv\xi \Lambda$ and $\textbf{Y}$ is a 3-dimensional vector setting the Dirac mass scale for active neutrinos through $v\textbf{Y}=m_D$. 
Notice that the ordering of the second and third column/row  of Eq.~(\ref{eq:MLISS}) is due to the assignment $L=+1$ and $-1$, for $N_1$ and $N_2$, respectively. Note also that 
the vanishing $(4,4)$ entry of the matrix in Eq.~(\ref{eq:MLISS}) would correspond to an additional LNV violation by two units, which does not generate neutrino masses at tree level but does so  only at loop level~\cite{Dev:2012sg,Lopez-Pavon:2012yda}. These loop corrections will be relevant only  for regimes of a large lepton number violation, which is not the case of this mechanism based on small lepton number violation.

In the seesaw limit, where $\left|m_D \right|, \left|\epsilon \right|, |\mu| \ll \Lambda$, we can perform the following block diagonalization: 
\begin{eqnarray}
U_B^T M_{\rm LISS} U_B = 
\left(
\begin{matrix}
m_{\text{light}}^{3\times 3} & \begin{matrix}
0_{3\times1 } & 0_{3\times1 }
\end{matrix} \\
\begin{matrix} 0_{1 \times 3} \\ 0_{1\times 3 } \end{matrix} & M_{\text{heavy}}^{2\times 2}
\end{matrix}
\right)\ ,
\label{eq:LISS_block_diagonalized}
\end{eqnarray}
where $U_B$ is a unitary matrix, and $m_{\text{light}}$ and $M_{\text{heavy}}$ are the mass matrices of the light and heavy sectors, respectively, given by:
\begin{eqnarray}\label{eq:LISS-mlight}
&&m_\nu\equiv m_{\text{light}} \simeq \frac{1}{\Lambda} \left( \mu \frac{Y_N Y_N^T v^2}{\Lambda} - \left( \epsilon v^2 Y_N {Y_N^\prime}^T + \epsilon v^2 Y_N^\prime Y_N^T \right) \right)
\ ,\\
&&M_{\text{heavy}} \simeq 
\left(
\begin{matrix}
0 & \Lambda  \\
\Lambda & \mu
\end{matrix}
\right)\ .
\label{eq:Mheavy}
\end{eqnarray}
Identifying 
$Y_N^{\prime\prime} \to Y_N^{\prime} + \frac{\mu v}{2   \epsilon \Lambda} Y_N$ in Eq.~(\ref{eq:LISS-mlight}), the light neutrino mass $m_{\nu}$ can be rewritten in a more concise form as:
\begin{equation}
m_{\nu} = - \frac{\epsilon v^2 Y_N Y_N'^T + \epsilon v^2 Y_N' Y_N^T}{\Lambda}\ .
\label{eq:LISS_mlight_rewriten}
\end{equation}

\subsection{Scan of the parameter space}
\label{subsec:scan}

Having established the theoretical framework of the LISS model in Subsection~\ref{subsec:model_LISS}, we now investigate which regions of its parameter space are consistent with current neutrino oscillation data and could produce observable signatures at DUNE. The challenge lies in efficiently exploring the multi-dimensional parameter space while ensuring that the model reproduces the observed neutrino mass-squared differences and mixing angles. A direct parameter scan starting from Eq.~\eqref{eq:MLISS} to satisfy neutrino data would be numerically inefficient. Therefore, following the same philosophy as in the Casas-Ibarra parametrization~\cite{Casas:2001sr}, we will express the model parameters explicitly in terms of the leptonic mixing angles and the light neutrino masses.

We begin by rewriting the Yukawa coupling vectors $\mathbf{Y}$ and $\mathbf{Y}'$ as:
\begin{equation}
    \mathbf{Y}= y \mathbf{u}, \quad \mathbf{Y}'= y' \mathbf{v},
    \label{rewriting_YYp}
\end{equation}
where $y$ and $y'$ are real numbers, $\mathbf{u}$ and $\mathbf{v}$ are normalized vectors, and $\mathbf{u}\cdot\mathbf{v}=\rho$. The light neutrino mass matrix is diagonalized as $m_{\nu} = \mathbf{U}^* \text{Diag}\{m_1, m_2, m_3 \} \mathbf{U}^{\dagger}$ (with $\mathbf{U}\equiv U_\text{PMNS}$).
 
This allows us to express the Yukawa couplings $\mathbf{Y},\, \mathbf{Y'}$ explicitly in terms of the leptonic mixing matrix elements~\cite{Gavela:2009cd}:

\begin{align}
Y_{j}^\text{NO} &= \frac{y}{\sqrt{2}} \left( \sqrt{1+\rho}\,\textbf{U}^*_{j3} + \sqrt{1-\rho}\,\textbf{U}^*_{j2} \right), \nonumber\\
Y_{j}^{\prime\,\text{NO}} &= \frac{y'}{\sqrt{2}} \left( \sqrt{1+\rho}\,\textbf{U}^*_{j3} - \sqrt{1-\rho}\,\textbf{U}^*_{j2} \right) - \frac{\mu}{2 \epsilon\Lambda} Y_{j}^{\text{NO}},
\label{eq:LISS_mD_NO} \\[0.3em]
Y_{j}^\text{IO} &= \frac{y}{\sqrt{2}} \left( \sqrt{1+\rho}\,\textbf{U}^*_{j2} + \sqrt{1-\rho}\,\textbf{U}^*_{j1} \right), \nonumber\\
Y_{j}^{\prime\,\text{IO}} &= \frac{y'}{\sqrt{2}} \left( \sqrt{1+\rho}\,\textbf{U}^*_{j2} - \sqrt{1-\rho}\,\textbf{U}^*_{j1} \right) - \frac{\mu}{2 \epsilon\Lambda} Y_{j}^{\text{IO}}.
\label{eq:LISS_mD_IO}
\end{align}
where $\rho$ depends on the hierarchy in the light neutrino spectrum: for normal ordering (NO), $\rho = 1-\frac{2|m_2|}{|m_3|+|m_2|}$, and for inverted ordering (IO), $\rho = 1-\frac{2|m_1|}{|m_1|+|m_2|}$. The light neutrino mass spectrum is then given by\footnote{Since the mass matrix in Eq.~\eqref{eq:MLISS} has rank at most 4, only four neutrinos can be massive, rendering the lightest neutrino massless.}:
\hspace*{-1cm}\begin{eqnarray}
\hskip -0.5cm \mathrm{NO}:\,\,\,&m_1=0,\,\,\,|m_2|=\frac{\epsilon y y' v^2}{\Lambda}(1-\rho),\,\,\,|m_3|=\frac{\epsilon y y' v^2}{\Lambda}(1+\rho)
\label{eq:mass_neutrino_light1}
\\
\mathrm{IO}:\,\,\,&m_3=0,\,\,\,|m_1|=\frac{\epsilon y y' v^2}{\Lambda}(1-\rho),\,\,\,|m_2|=\frac{\epsilon y y' v^2}{\Lambda}(1+\rho)
\label{eq:mass_neutrino_light2}
\end{eqnarray}
From Eqs.~\eqref{eq:mass_neutrino_light1} and~\eqref{eq:mass_neutrino_light2}, it is possible to constrain the product $\epsilon y'$ for each mass ordering as:
\hspace*{-1cm}\begin{eqnarray}
\hskip -0.5cm \mathrm{NO}:\,\,\,&\epsilon y'= \frac{\Lambda(|m_2|+|m_3|)}{2 y v^2}\\
\mathrm{IO}:\,\,\,&\epsilon y'= -\frac{\Lambda(|m_1|+|m_2|)}{2 y v^2}
\label{eq:constrain_ypepsilon}
\end{eqnarray}
Finally, the eigenstates of the heavy mass matrix in Eq.~\eqref{eq:Mheavy} are given by:
\begin{equation}
m_{4,5} \simeq \Lambda \pm \frac{1}{2} |\mu| \ . 
\end{equation}

We are now ready to perform our parameter scan. Equations~\eqref{eq:LISS_mD_NO} and~\eqref{eq:LISS_mD_IO} were derived in the seesaw limit, yet some viable parameter space regions that reproduce neutrino oscillation data might deviate from this limit. To account for these possibilities while keeping the scan numerically feasible, we introduce perturbations to our equations as follows:
\begin{align}
Y_{j,\text{scan}}^{\rm NO} &= \bigl(1+\delta_j^{\rm NO}\bigr)\,Y_j^{\rm NO}
\;;\quad
Y_{j,\text{scan}}^{\prime\,\rm NO} = \bigl(1+\delta_j^{\prime\,\rm NO}\bigr)\,Y_j^{\prime\,\rm NO}
\label{eq:LISS_scan_NO},\\
Y_{j,\text{scan}}^{\rm IO} &= \bigl(1+\delta_j^{\rm IO}\bigr)\,Y_j^{\rm IO}
\;;\quad
Y_{j,\text{scan}}^{\prime\,\rm IO} = \bigl(1+\delta_j^{\prime\,\rm IO}\bigr)\,Y_j^{\prime\,\rm IO}
\label{eq:LISS_scan_IO}\ ,
\end{align}
\cm{where we let $\delta_j^{\rm IO},\delta_j^{\rm NO}$ vary between 0 and 1, keeping only the points that reproduce neutrino oscillation data (see Appendix~\ref{App:neutrino_oscillation_data}). In our 100-million-point scan, perturbations up to 20\% appeared sufficient, as larger variations did not indicate any noticeable changes in the allowed parameter space.}

\begin{figure}
  \centering
  \includegraphics[width=0.5\textwidth]{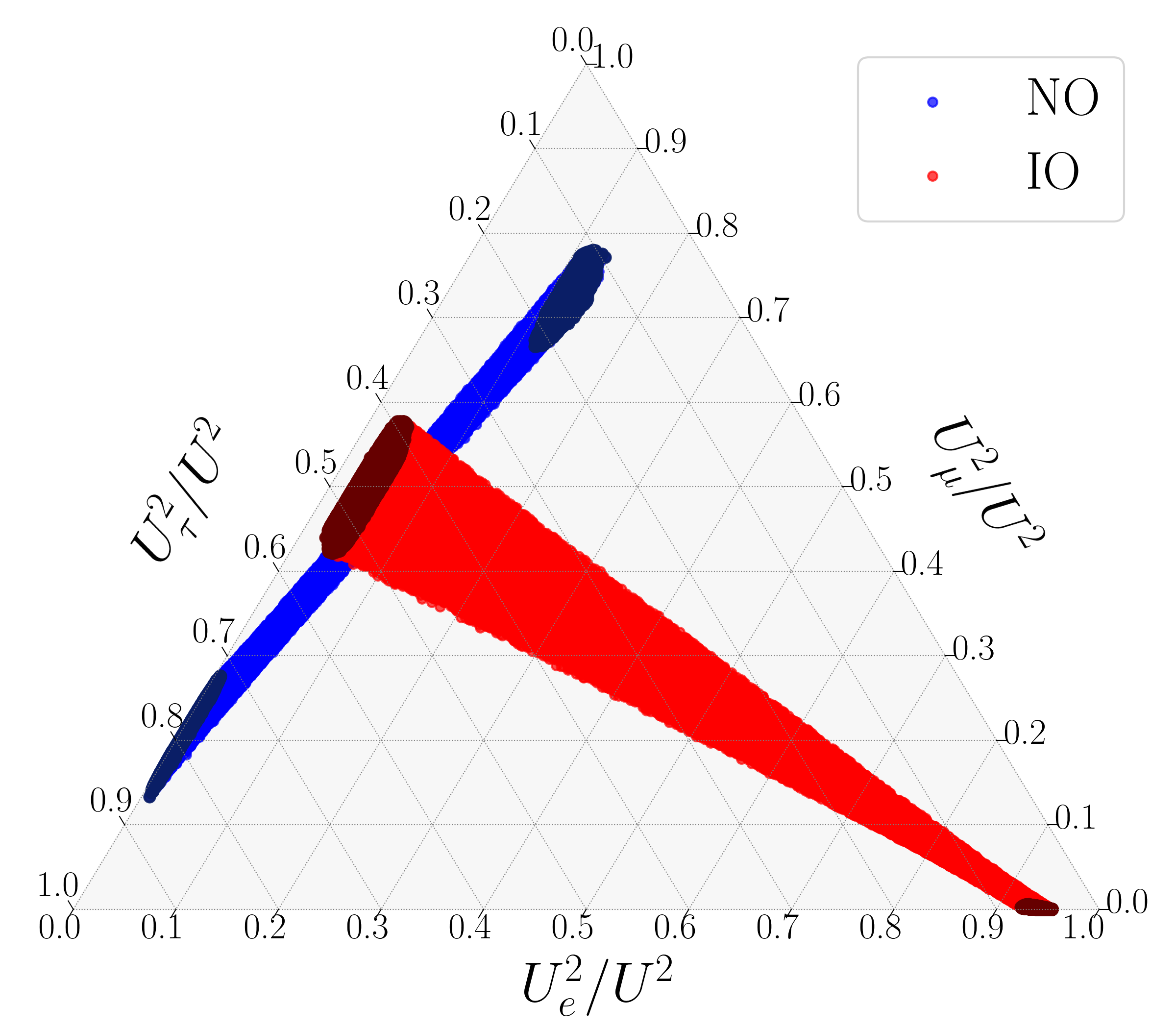}
  \caption{Allowed flavor structures in the LISS model consistent with neutrino oscillation data are shown for normal ordering (blue) and inverted ordering (red). \cm{Points with $U^4 > 1 \times 10^{-6}$ are highlighted in darker colors}. Here, $U^2_{\alpha} \equiv U^2_{\alpha4} + U^2_{\alpha5}$ and $U^2 \equiv \sum_{\alpha} U^2_{\alpha}$.
}.
   \label{fig:ternary}
\end{figure}

After performing the scan, we illustrate in Fig.~\ref{fig:ternary} the allowed flavor structures within the LISS model, \cm{expressed in terms of the normalized flavor ratios $U^2_{\alpha}/U^2$, where $U^2_{\alpha}\equiv U^2_{\alpha4}+U^2_{\alpha5}$ and $U^2\equiv \sum_{\alpha} U^2_{\alpha}$}. For NO, $m_3\gg m_2$, thus the flavor structure is dominated by the third column of the leptonic mixing matrix, suppressing electron flavor contributions due to the smallness of $\theta_{13}$. In contrast, IO features $m_2\approx m_1$, allowing contributions from both the first and second columns of the leptonic mixing matrix, yielding a broader range of flavor structures, i.e.,   from electron-dominated to more democratic textures.

\cm{It is important to note that, although all the flavor structures in the allowed regions of Fig.~\ref{fig:ternary} are consistent with neutrino data, from a phenomenological point of view the most accessible flavour structures are those shown in dark, corresponding to points with $U^4 > 1 \times 10^{-6}$. These coincide with the extremes of the allowed regions.
}

For simplicity, current bounds and future sensitivities often consider only single-flavor mixing scenarios~\cite{Antel:2023hkf}. However, such structures generally do not reproduce neutrino oscillation data. It would thus be preferable to present current bounds and future sensitivities using more realistic flavor structures as suggested in Refs.~\cite{Abada:2018sfh,Abada:2022wvh}, and such as the two benchmarks proposed in Ref.~\cite{Drewes:2022akb}:

\begin{equation}
    U_e^2: U_\mu^2: U_\tau^2=0: 1: 1
\end{equation}
and 
\begin{equation}
    U_e^2: U_\mu^2: U_\tau^2=1: 1: 1,
\end{equation}
which fall within the regions allowed by neutrino oscillation data.

\section{Experimental set-up}
\label{sec:exp_setup}

To evaluate DUNE's capability to probe the 3+(pseudo-Dirac pair) oscillation framework, as well as the LISS model developed in the previous sections, we must translate the theoretical predictions into experimentally observable signatures. The quasi-degenerate sterile neutrino masses ($\sim$keV) and the specific flavor mixing patterns predicted by our setup require careful consideration of detector baselines, energy resolutions, and systematic uncertainties. This section outlines the experimental configuration and simulation methodology used to assess DUNE's sensitivity to these BSM scenarios.

DUNE represents the next generation of long-baseline neutrino oscillation experiments and is currently under construction in the United States. Upon completion, the experiment will feature two state-of-the-art neutrino detector complexes (near and far detectors) exposed to a very intense neutrino beam, the Long-Baseline Neutrino Facility (LBNF) beam, generated by 120 GeV protons \cm{with a beam power of 1.2~MW~\cite{DUNE:2020lwj}}. This neutrino beam peaks in energy between 2 and 3 GeV. The Near Detector (ND) complex will be situated at Fermilab in Illinois, close to where neutrinos are produced, whereas the significantly larger Far Detector (FD) will be positioned approximately 1.5 km underground at the Sanford Underground Research Facility (SURF) in South Dakota, around 1300 km away.

The ND complex, located 574 meters downstream from the neutrino beam source, will include three main detectors. The first one is an advanced Liquid Argon Time Projection Chamber (LArTPC), known as ND-LAr, \cm{with a fiducial mass of about 67 tons~\cite{DUNE:2021tad}}. The second detector is the System for on-Axis Neutrino Detection (SAND), and the third one is the Muon Spectrometer (TMS), which will later be upgraded to a gaseous argon TPC (ND-GAr). On the other hand, during the experiment’s first phase, the FD complex will consist of two large LArTPC modules utilizing both vertical and horizontal drift, providing a total fiducial volume of 40 kilotons~\cite{DUNE:2020lwj}.

\subsection{Simulation details}
\label{subsec:Simulation_details}
In our analysis, we simulate the expected event rates at DUNE’s near and far LArTPC detectors, using configurations outlined in the DUNE Technical Design Report (TDR) (see Refs. \cite{DUNE:2020ypp,DUNE:2020jqi}). Specifically, we use the nominal beam setup and assume data collection of 3.5 years in neutrino mode and another 3.5 years in antineutrino mode. All simulations are conducted with the GLoBES software~\cite{Huber:2004ka,Huber:2007ji}, using the auxiliary files from the DUNE collaboration (see Ref.~\cite{DUNE:2021cuw}), where we take their systematic uncertainties, which range between 5\% and 20\%, depending on the specific channel. It should be noted that the systematics included in the auxiliary files of GLoBES provided by the DUNE collaboration are purely normalization uncertainties (i.e. global normalization systematics). However, there are also systematics arising from effects such as beam focusing and detailed cross-section modeling (including nuclear effects in quasi-elastic scattering (QE) and resonant scattering (RES) processes) that distort the expected energy spectrum. 

These are commonly referred to in the literature as “shape uncertainties,” and their importance has been demonstrated both for the determination of standard oscillation parameters \cite{Coyle:2025xjk} and for BSM searches \cite{DUNE:2021tad,Miranda:2018yym,Coloma:2021uhq,Meloni:2018xnk}. In what follows, we assume that these shape uncertainties can be kept under control and will not significantly alter the conclusions drawn from our analysis.  Their full treatment lies beyond the scope of the present  work.

\subsection{\texorpdfstring{$\tau$}{tau} detection}
\label{subsec:tau_detec}
The auxiliary files provided by the DUNE collaboration do not include any explicit $\tau$ detection. For our simulation of $\nu_\tau$-like events, relevant for channels involving tau neutrino appearance,  we use the $\nu_\tau$ charged-current cross section on argon given in the ancillary files of Ref.~\cite{Alion:2016uaj}, generated with GENIE v2.8.4~\cite{Andreopoulos:2015wxa,Tena-Vidal:2021rpu} and we adopt a methodology for $\tau$ detection similar to the one described in Ref.~\cite{Coloma:2021uhq,DeGouvea:2019kea}, which we summarize in the remainder of the section. 

A $\tau$ produced at a few GeV travels only millimeters before decaying, far too short to yield a visible track in liquid argon. What remains is either a hadronic cascade or a charged lepton plus missing energy. Since hadronic modes account for roughly 65\% of all $\tau$ decays, while each leptonic mode contributes only about 17\%~\cite{Zyla:2020zbs}, and due to the fact that the leptonic channels suffer greater contamination from charged current (CC) $\nu_e$ and $\nu_\mu$ interactions, we conservatively consider only the hadronic $\tau$ decays. We assume a signal‐detection efficiency of 30\% for these modes. The energy smearing for $\nu_\tau$ events is modeled by assuming the observed energy follows a Gaussian distribution with a mean of $0.45 E_\nu^{\text{true}}$ and a width of $0.25 E_\nu^{\text{true}}$, where $E_\nu^{\text{true}}$ is the true incident neutrino energy. The primary background for hadronic $\tau$ decays comes from neutral current (NC) interactions. For this NC background, we use the migration matrices provided by the DUNE collaboration for $\nu_e$ CC events, and apply a constant background rejection efficiency of $0.5\%$, as suggested in Ref.~\cite{DeGouvea:2019kea}.  Additionally, we consider a 20\% normalization systematic uncertainty for all channels involving tau detection.

\section{Results}
\label{sec:results}

In this section, we present DUNE's sensitivity to the 3+(pseudo-Dirac pair) oscillation scenario and its implications for the LISS model. Building on the theoretical framework developed in Sections~\ref{sec:oscillation_LISS} and~\ref{sec:LISS}, and using the experimental simulation approach described in Section~\ref{sec:exp_setup}, we evaluate DUNE's projected sensitivity for both disappearance and appearance channels, investigating the complementarity between the near and far detectors, first assuming no extra CP violation arising from the new physics scenario, and second evaluating the impact of the new leptonic CP-violating phases on the sensitivity. All our results are presented at the 90$\%$ C.L. using the frequentist approach\footnote{Wilks’ theorem~\cite{Wilks:1938dza} is assumed to be valid.} 
($\Delta\chi^2 > 4.61$), assuming 3.5 years of data collection in both neutrino and antineutrino modes.

\begin{figure}[!htb]
  \centering
  \begin{tabular}{cc}
  \includegraphics[width=0.45\textwidth]{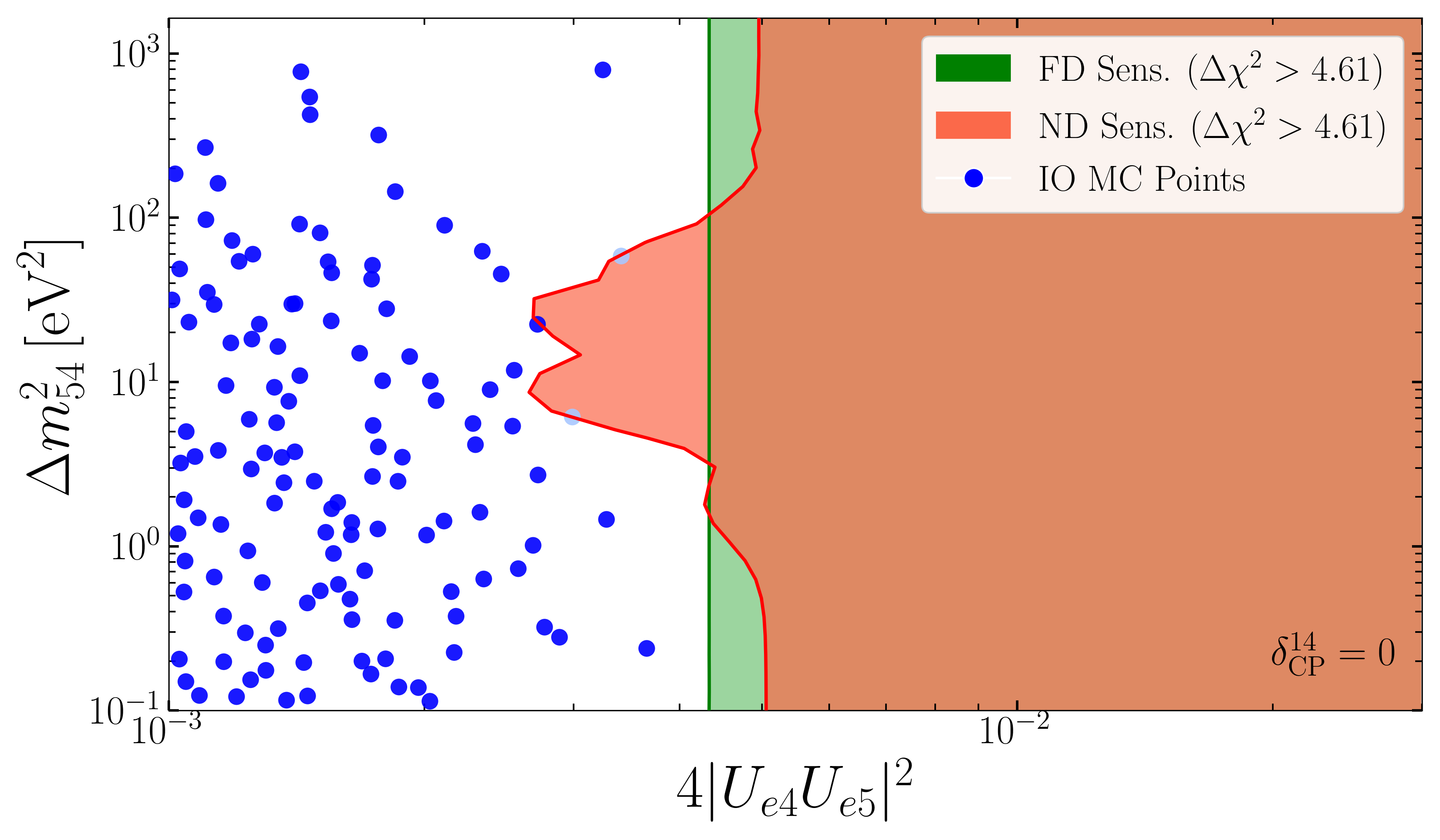}
  &\includegraphics[width=0.45\textwidth]{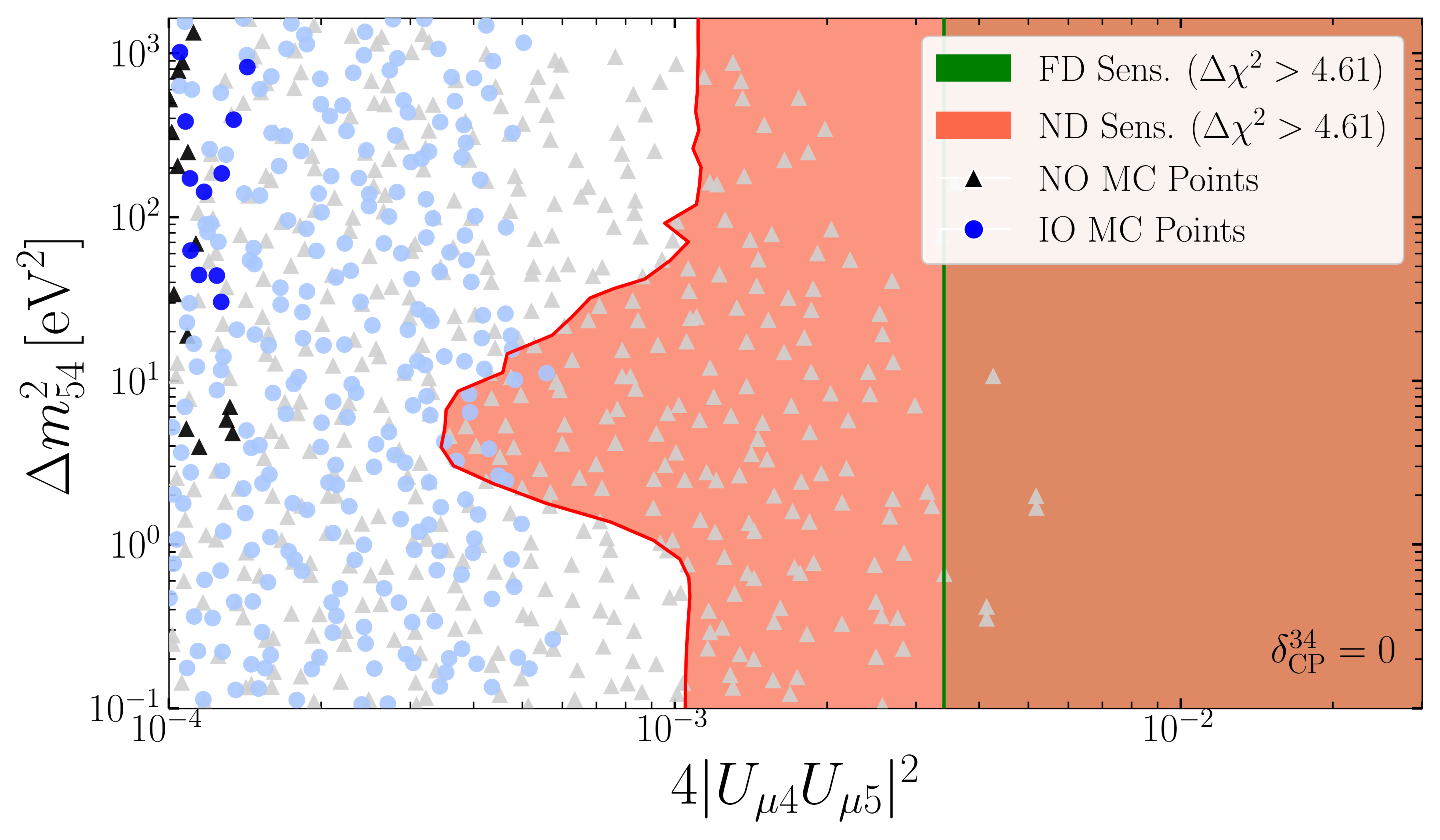}\\
  (A)&(B)\\
  \includegraphics[width=0.45\textwidth]{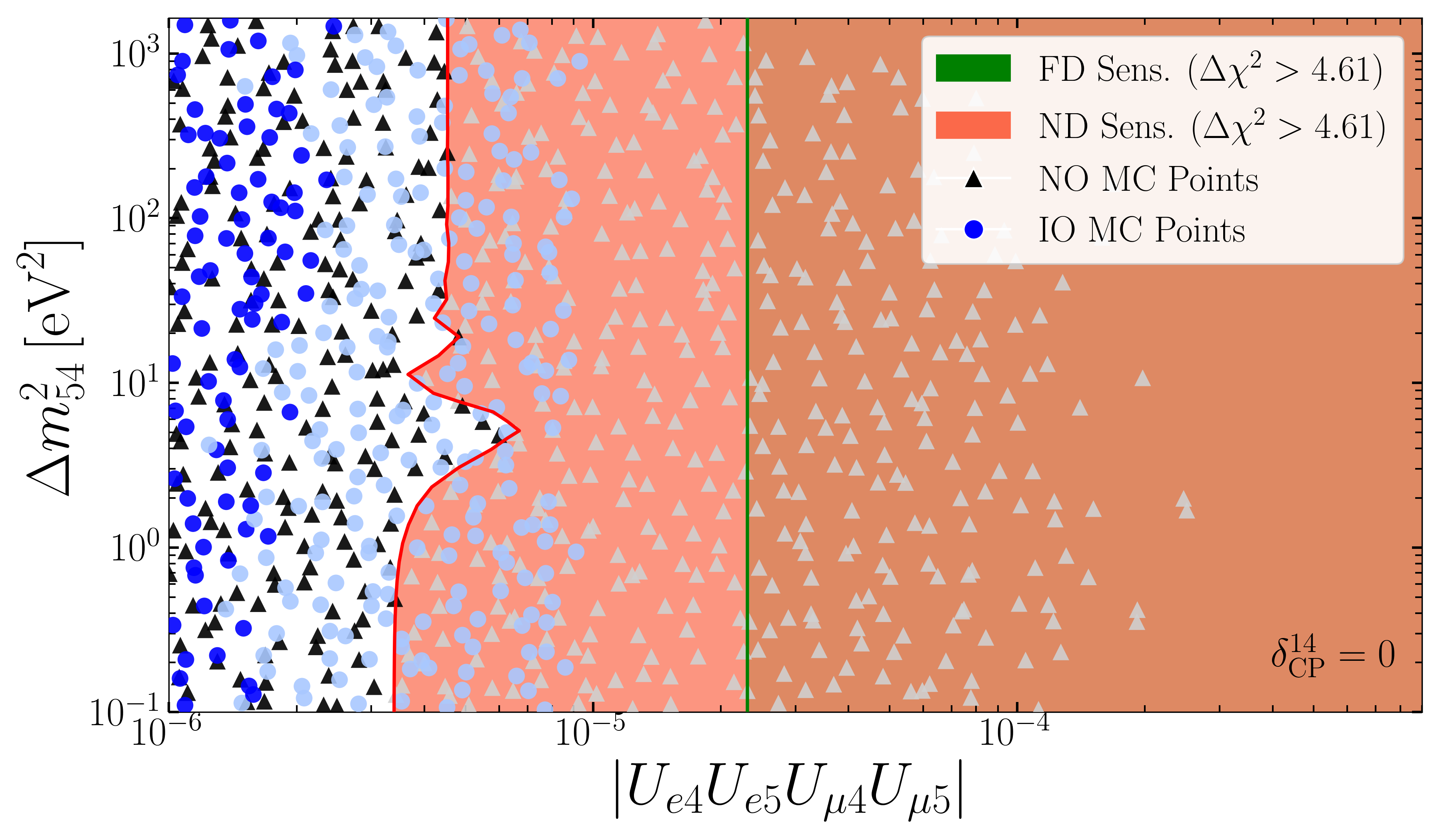}
  &\includegraphics[width=0.45\textwidth]{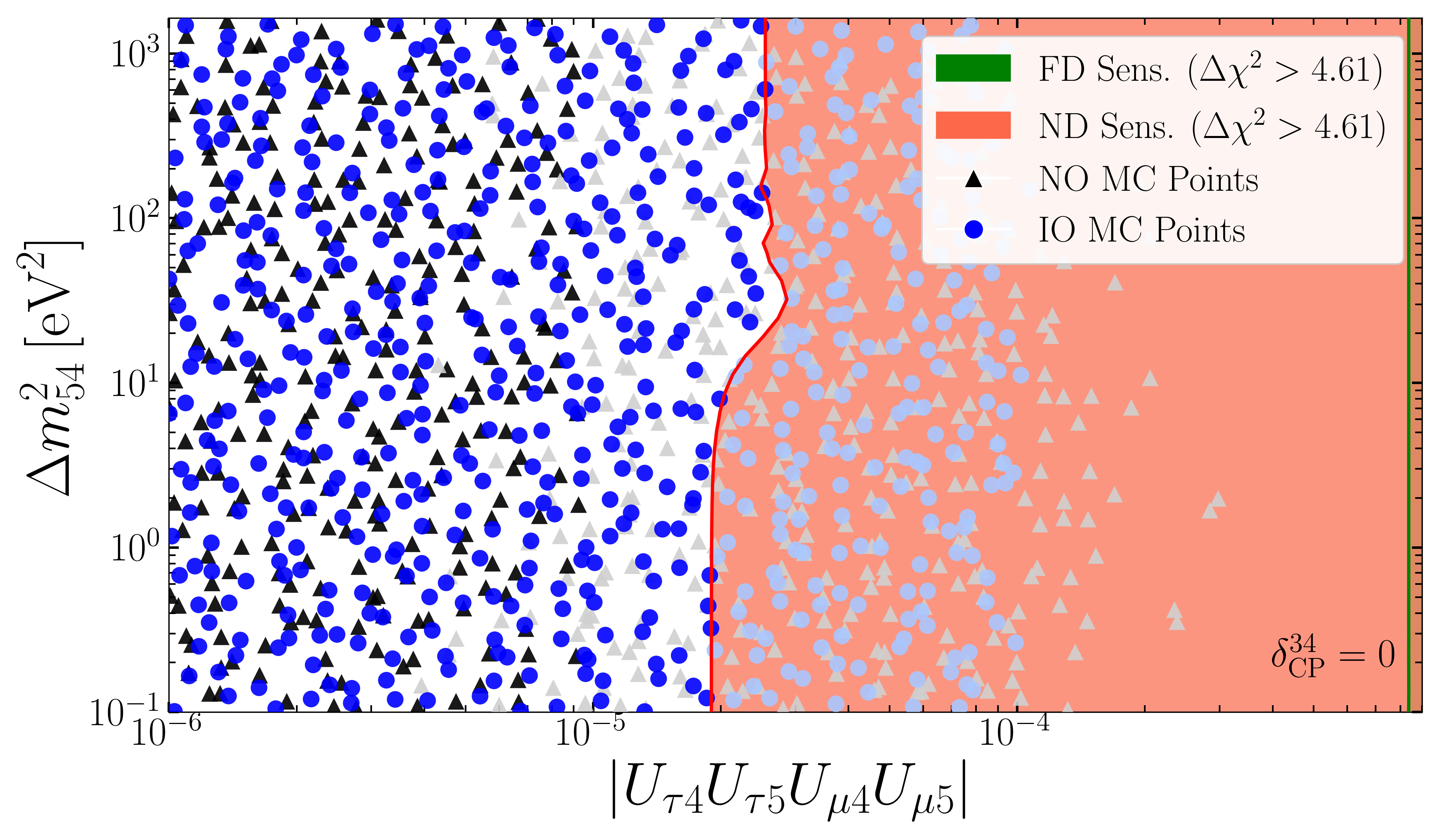}\\
  (C)&(D)
   \end{tabular}
  \caption{DUNE sensitivity to 3+pseudo-Dirac neutrino oscillations across all channels. \\
  \textbf{(A):} Electron neutrino disappearance sensitivity. \textbf{(B):} Muon neutrino 
  disappearance sensitivity. \textbf{(C):} 
  Electron neutrino appearance sensitivity probing $|U_{e4}U_{e5}U_{\mu4}U_{\mu5}|$. 
  \textbf{(D):} and the analogous ones for tau neutrino appearance. 
  Red contours show 90$\%$ C.L. exclusion regions for the Near Detector, while green contours 
  correspond to the Far Detector. Blue dots represent allowed LISS model points for Inverted 
  Ordering, and black triangles show Normal Ordering predictions. \cm{The points that will be probed by DUNE are shown in lighter colors}. All results assume vanishing 
  new CP phases ($\delta_{\text{CP}}^{ij} = 0$).}
  \label{fig:appdis}
\end{figure}

Figure~\ref{fig:appdis} presents DUNE's sensitivity to (pseudo-Dirac pair) sterile neutrino oscillations \cm{for the relevant combination of the mixing parameters entering in Eq.~\eqref{eq:app_prob} and Eq.~\eqref{eq:dis_prob}, assuming no CP violation arising from new physics. The combinations $4\left|U_{\alpha 4} U_{\alpha 5}\right|^2$ are probed by the disappearance channels, whereas $\left|U_{\beta 4} U_{\beta 5} U_{\mu 4} U_{\mu 5}\right|$ with $\beta = e,\tau$ are probed by the appearance channels}\footnote{\cm{In principle, the combination $|U_{e4} U_{e5} U_{\mu4} U_{\mu5}|$ is also constrained by disappearance channels. However, we find that the appearance channel provides the best sensitivity.}}. To ease the comparison with current bounds, we have also generated Table~\ref{tab:constraints_summary}, in which we show the current limits and the \cm{DUNE expected sensitivity} for the three different mass regimes introduced at the end of Section~\ref{sec:oscillation_LISS}. \cm{\cm{A detailed explanation of the method used to obtain them from the $3+1$ scenario results is provided in Appendix~\ref{App:Constraints}.}
}

\begin{table}[h!]
\centering
\renewcommand{\arraystretch}{1.1}
\begin{tabular}{|l|l|c|c|}
\hline
\textbf{\cm{Parameter}} & \textbf{Channel} & \textbf{Current ($90\%$ C.L.) } & \textbf{DUNE expected ($90\%$ C.L.) } \\
\hline
$4|U_{e4} U_{e5}|^2$ & $\nu_e$ DIS & 
$\begin{cases}
0.00036  & \text{LM \cite{Goldhagen:2021kxe}} \\
0.00036  & \text{R \cite{Goldhagen:2021kxe}} \\
0.00036  & \text{HM \cite{Goldhagen:2021kxe}}
\end{cases}$ &
$\begin{cases}
0.0043  & \text{LM \cite{Goldhagen:2021kxe}} \\
0.0027  & \text{R \cite{Goldhagen:2021kxe}} \\
0.0043  & \text{HM \cite{Goldhagen:2021kxe}}
\end{cases}$ \\
\hline
$4|U_{\mu4} U_{\mu5}|^2$ & $\nu_\mu$ DIS &
$\begin{cases}
0.00076  & \text{LM \cite{MINOS:2020iqj}} \\
0.00076  & \text{R \cite{MINOS:2020iqj}} \\
0.00076  & \text{HM \cite{MINOS:2020iqj}}
\end{cases}$ &
$\begin{cases}
0.0011  & \text{LM } \\
\textcolor{blue}{0.00035} & \text{R } \\
0.0011  & \text{HM }
\end{cases}$ \\
\hline
$|U_{e4} U_{e5} U_{\mu4} U_{\mu5}|$ & $\nu_e$ APP &
$\begin{cases}
0.00011 & \text{LM \cite{NOMAD:2001xxt}} \\
<0.00014 & \text{R \cite{NOMAD:2001xxt}} \\
0.00014 & \text{HM \cite{NOMAD:2001xxt}}
\end{cases}$ &
$\begin{cases}
\textcolor{blue}{3.4 \cdot 10^{-6}} & \text{LM} \\
\textcolor{blue}{8.2 \cdot 10^{-6}} & \text{R } \\
\textcolor{blue}{4.6 \cdot 10^{-6}} & \text{HM}
\end{cases}$ \\
\hline
$|U_{\tau4} U_{\tau5} U_{\mu4} U_{\mu5}|$ & $\nu_\tau$ APP &
$\begin{cases}
2.7 \cdot 10^{-5} & \text{LM \cite{NOMAD:2001xxt}} \\
 <3.6 \cdot 10^{-5}& \text{R}~\cite{NOMAD:2001xxt} \\
3.6 \cdot 10^{-5} & \text{HM \cite{NOMAD:2001xxt}}
\end{cases}$ &
$\begin{cases}
\textcolor{blue}{1.9 \cdot 10^{-5}} & \text{LM} \\
\textcolor{blue}{3.0 \cdot 10^{-5}} & \text{R} \\
\textcolor{blue}{2.5 \cdot 10^{-5}} & \text{HM}
\end{cases}$ \\
\hline
\end{tabular}
\caption{Current constraints and projected DUNE sensitivity for the $3+$(pseudo-Dirac pair) model in the low-mass (LM), resonant (R), and high-mass (HM) regimes. In the resonant regime, we show the limit at the peak of the resonance for disappearance (DIS) channels. For appearance (APP) channels, \cm{the resonant regime, due to destructive interference, always yields a sensitivity worse than in the HM regime}. We highlight in blue when DUNE is expected to improve over current constraints.}
\label{tab:constraints_summary}
\end{table}

Panel (A) of Figure~\ref{fig:appdis} shows the sensitivity to electron neutrino disappearance 
through the parameter $4|U_{e4}U_{e5}|^2$. The FD provides stronger bounds than the ND for 
$\Delta m_{54}^2>100\,\text{eV}^2$ and $\Delta m_{54}^2<1\,\text{eV}^2$, with limits of 
$4|U_{e4}U_{e5}|^2 < 4.3\times 10^{-3}$ due to the interference with the standard oscillation with three flavors, which are negligible for the ND but become dominant for the FD. In the intermediate range 
$1\,\text{eV}^2<\Delta m_{54}^2<100\,\text{eV}^2$, the ND dominates with significantly 
stronger constraints reaching $4|U_{e4}U_{e5}|^2 < 2.7\times 10^{-3}$, as the ND baseline allows for a resonant behavior which improves the sensitivity. However, and as shown in Table~\ref{tab:constraints_summary}, these sensitivities are still weaker than the current bound arising from solar neutrinos~\cite{Goldhagen:2021kxe}, even in the resonant regime. This is due to the fact that, as can be seen in Eq.~\eqref{eq:dis_prob}, the resonant contribution to the disappearance probability only enters at second order, whereas the non-unitarity contribution already appears at leading order in the mixing parameters. This implies that the expected improvement for the resonant regime is less in disappearance than in other sterile scenarios like the $3+1$ one.

Panel (B) of Figure~\ref{fig:appdis} shows that muon neutrino disappearance provides a stronger 
constraint than electron neutrino disappearance. The ND constraint is better than the FD across 
all values of $\Delta m_{54}^2$, with sensitivity limits of 
$4|U_{\mu4}U_{\mu5}|^2<3.5\times 10^{-4}$ in the resonant regime 
($1\,\text{eV}^2<\Delta m_{54}^2<70\,\text{eV}^2$). Outside this range, 
the limits relax to $4|U_{\mu4}U_{\mu5}|^2<1.05\times 10^{-3}$. As shown  in Table~\ref{tab:constraints_summary}, 
this constitutes an improvement over existing bounds from MINOS~\cite{MINOS:2020iqj} in the resonance region.

The ND dominance in muon disappearance probability  mainly arises from two factors: (i) the high 
initial $\nu_\mu$ flux in the DUNE beam, providing excellent statistics, and (ii) the 
short baseline that probes the sterile oscillation frequency. Conversely, 
the FD observes a significantly reduced $\nu_\mu$ flux, substantially reducing 
its capabilities for sterile neutrino searches via $\nu_\mu$ disappearance.

Panels (C) and (D) of Figure~\ref{fig:appdis} show the sensitivity for appearance channels. We present our results in the parameter combinations 
$|U_{\ell 4}U_{\ell 5}U_{\mu 4}U_{\mu 5}|$ for $\ell = e,\,\tau$. Panel (C) of Figure~\ref{fig:appdis} exhibits a much better sensitivity for the 
ND than for the FD.
In the limit of high values of $\Delta m_{54}^2$, the ND achieves 
$|U_{e 4}U_{e 5}U_{\mu 4}U_{\mu 5}|<4.6\times 10^{-6}$, while the FD reaches 
$|U_{e 4}U_{e 5}U_{\mu 4}U_{\mu 5}|<2.3\times 10^{-4}$. Similarly, for tau neutrino appearance (Panel (D) of Figure~\ref{fig:appdis}), 
the ND provides constraints of $|U_{\tau 4}U_{\tau 5}U_{\mu 4}U_{\mu 5}|<2.5\times 10^{-5}$, 
while the FD sensitivity (green contour) is weaker at 
$|U_{\tau 4}U_{\tau 5}U_{\mu 4}U_{\mu 5}|<7.5\times 10^{-4}$. In the limit of low values of $\Delta m_{54}^2$, the ND achieves 
$|U_{e 4}U_{e 5}U_{\mu 4}U_{\mu 5}|<3.4\times 10^{-6}$. Similarly, for tau neutrino appearance (Panel (D) of Figure~\ref{fig:appdis}), 
the ND provides constraints of $|U_{\tau 4}U_{\tau 5}U_{\mu 4}U_{\mu 5}|<1.9\times 10^{-5}$.
As detailed in Table~\ref{tab:constraints_summary}, these projected ND bounds improve significantly over current limits 
from NOMAD~\cite{NOMAD:2001xxt}, especially in the electron appearance channel, where the gain exceeds one order of magnitude.

Additionally, the dots and triangles throughout Figure~\ref{fig:appdis} represent viable parameter points 
for the LISS model after imposing constraints from standard three-neutrino oscillation data~\cite{Esteban:2024eli}. 
Blue dots correspond to IO, while black triangles represent NO. \cm{Points that will be probed by DUNE are shown in lighter colors}. An important feature is the complete absence of normal ordering solutions (black triangles) in the electron disappearance 
panel (A), this is due to the electron mixing suppression that we have for NO arising from the smallness of $\theta_{13}$ as explained at the end of Section~\ref{sec:LISS}. \cm{Once embedded in the LISS model, some interesting features appear. In the case of NO, the $\nu_e$ appearance channel provides the strongest sensitivity, whereas for IO it is the $\nu_{\tau}$ appearance channel. These two regions correspond in Fig.~\ref{fig:ternary} to the dark region around $U^2_{\mu}/U^2 = 0.7$, $U^2_{\tau}/U^2 = 0.2$, and $U^2_{e}/U^2 = 0.1$ for NO, and to the electron-suppressed region for IO. We observe that the points within reach of the $\nu_{\mu}$ disappearance channel will be probed by the appearance channels. On the other hand, the $\nu_e$ disappearance channel will still be able to probe part of the electron-dominated region of IO.
    }

\begin{figure}[!htb]
  \centering
\begin{tabular}{cc}
      \includegraphics[width=0.45\textwidth]{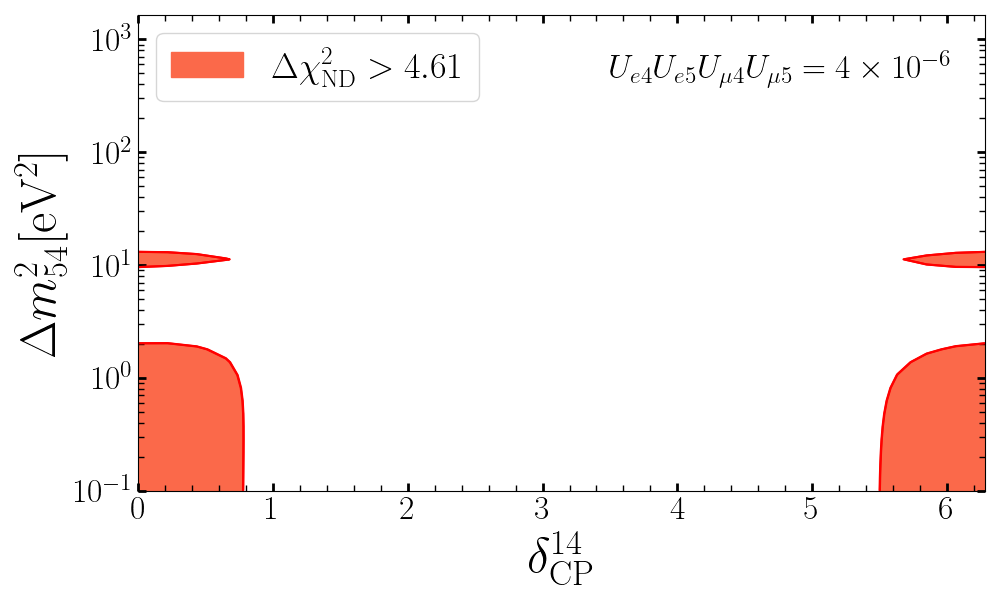}
  &\includegraphics[width=0.45\textwidth]{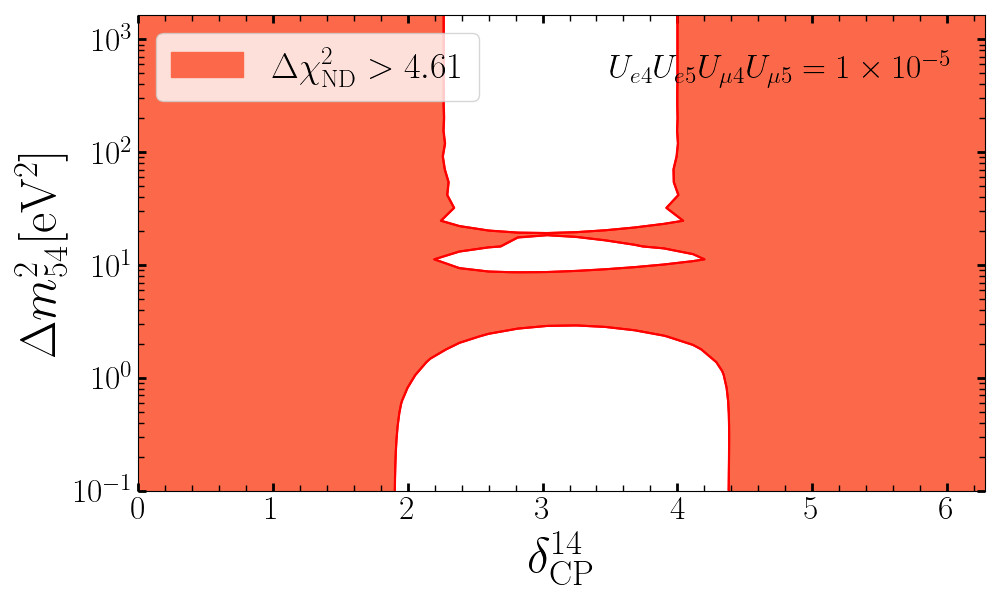}\\
  (A)&(B)\\
  \includegraphics[width=0.45\textwidth]{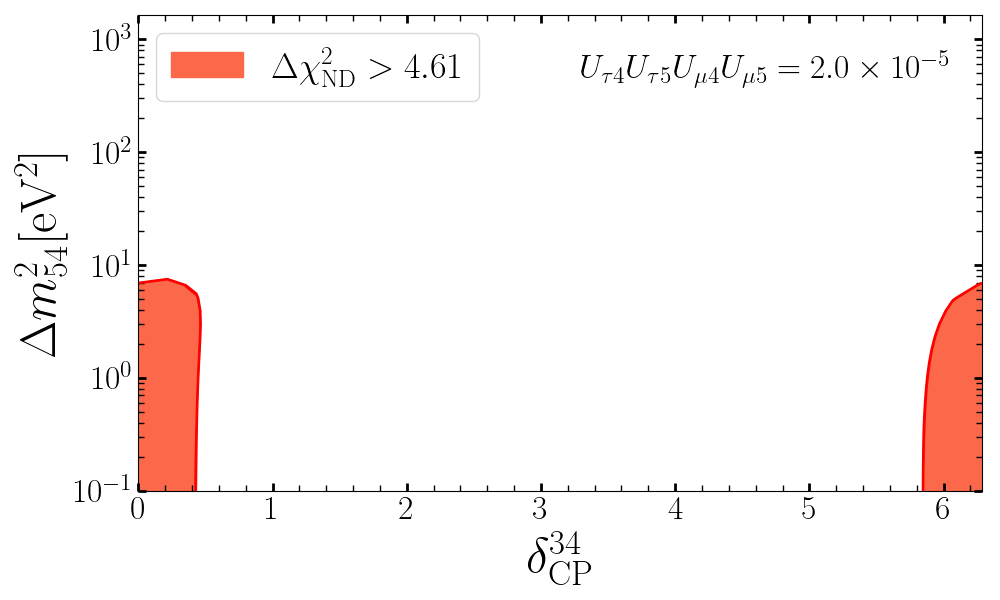}
  &\includegraphics[width=0.45\textwidth]{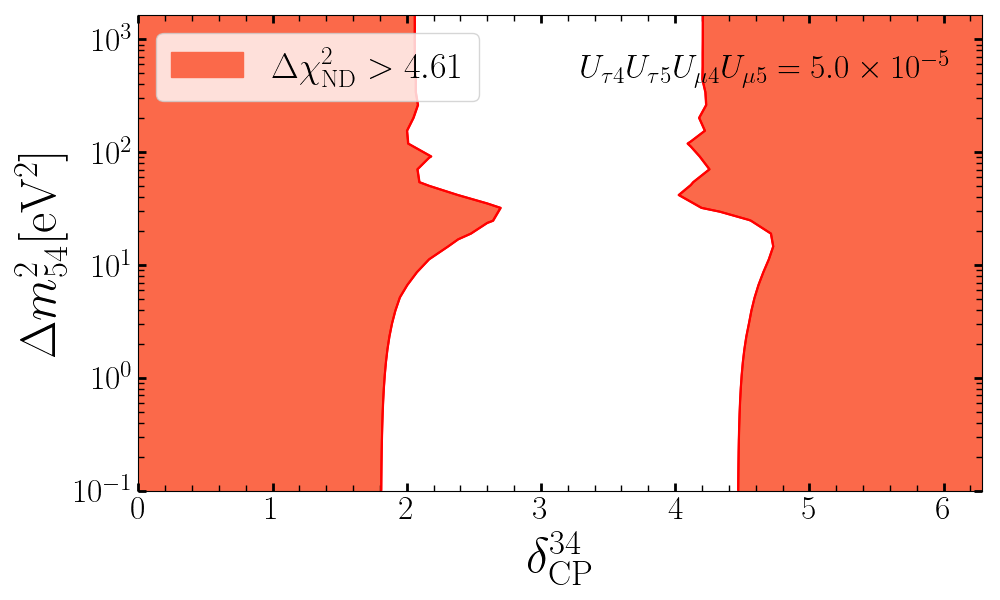}\\
    (C)&(D)
  \end{tabular}
  \caption{DUNE sensitivity to new leptonic CP-violating phases in sterile neutrino oscillations. 
  \textbf{(A) and (B):} Excluded regions in the $\Delta m_{54}^2$--$\delta_\text{CP}^{14}$ 
  plane for electron neutrino appearance, with fixed values of $|U_{e4}U_{e5}U_{\mu4}U_{\mu5}|$ 
  at $4.0\times10^{-6}$ (A) and $1.0\times10^{-5}$ (B). \textbf{ (C) and (D):} Sensitivity reach for 
  tau neutrino appearance in the $\Delta m_{54}^2$--$\delta_\text{CP}^{34}$ parameter space, 
  with fixed $|U_{\tau4}U_{\tau5}U_{\mu4}U_{\mu5}|$ values of $2.0\times10^{-5}$ (C) and $5.0\times10^{-5}$ (D). 
  The plots demonstrate how CP violating-phases can dramatically alter the experimental signature, 
  with sensitivity varying by orders of magnitude across the phase space. All results 
  correspond to 90$\%$ C.L. exclusion contours.}
  \label{fig:deltaCPnew}
\end{figure}
\begin{figure}[!htb]
  \centering
  \begin{tabular}{cc}
\includegraphics[width=0.45\textwidth]{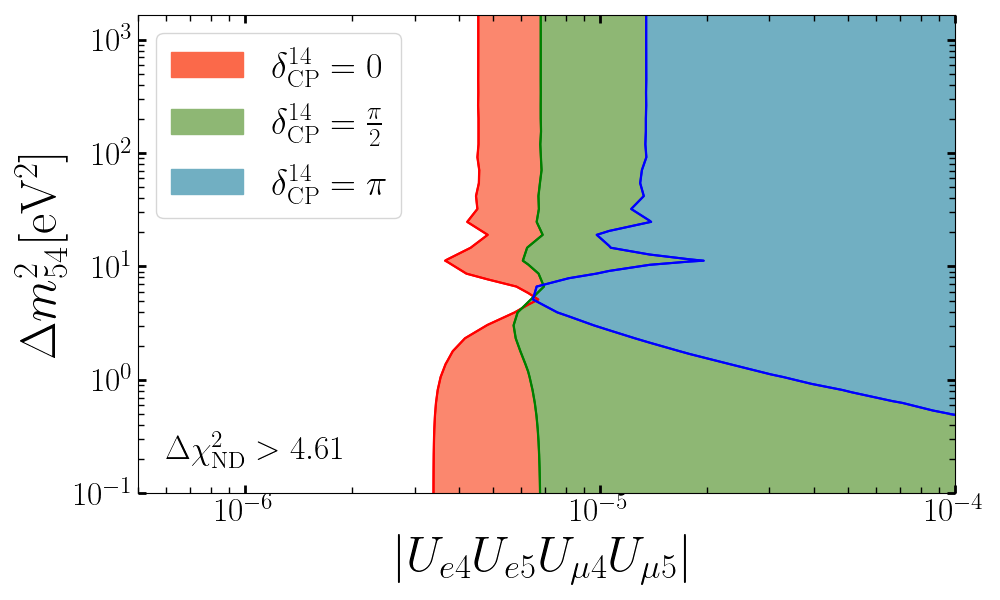}
&\includegraphics[width=0.45\textwidth]{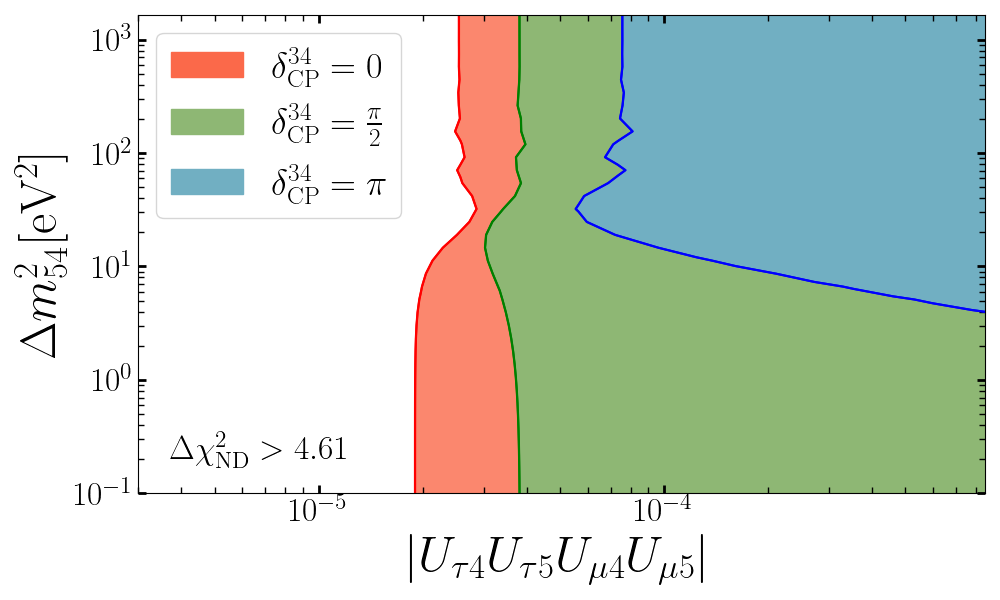}\\
(A)&(B)
\end{tabular}
  \caption{Impact of new leptonic CP-violating phases on DUNE appearance channel sensitivity. 
  \textbf{ (A):} Near Detector constraints on electron neutrino appearance 
  ($|U_{e4}U_{e5}U_{\mu4}U_{\mu5}|$ vs $\Delta m_{54}^2$) for three values of the 
  CP-violating phase $\delta_\text{CP}^{14}$: $0$ (red), $\pi/2$ (green), and $\pi$ (blue). 
  \textbf{ (B):} Corresponding sensitivity for tau neutrino appearance 
  ($|U_{\tau4}U_{\tau5}U_{\mu4}U_{\mu5}|$ vs $\Delta m_{54}^2$) with varying 
  $\delta_\text{CP}^{34}$ phases. The dramatic variation between red and blue contours 
  illustrates how CP-violating phases can either enhance or completely suppress the appearance 
  signal, with constructive interference at $\delta_\text{CP} = 0$ yielding to the best 
  sensitivity and destructive interference at $\delta_\text{CP} = \pi$ leading to 
  strong suppression. All exclusion regions correspond to 90$\%$ C.L.}
  \label{fig:appdeltaCPnew}
\end{figure}

The role of new leptonic CP-violating phases due to the presence of the sterile neutrino states is explored in 
Figures~\ref{fig:deltaCPnew} and~\ref{fig:appdeltaCPnew}. The expressions of the lepton  mixing matrix elements in terms of the mixing angles and of the CP-violating phases are presented in Appendix A. Figure~\ref{fig:deltaCPnew} demonstrates the sensitivity reach in the 
$\Delta m_{54}^2$ vs $\delta_\text{CP}^{14}$ parameter space for fixed values of the 
mixing parameters. Panels (A) and (B) focus on electron neutrino appearance with 
different fixed values of $|U_{e 4}U_{e 5}U_{\mu 4}U_{\mu 5}|$, while Panels (C) 
and (D) examine tau neutrino appearance with varying $|U_{\tau 4}U_{\tau 5}U_{\mu 4}U_{\mu 5}|$.

The complementary perspective is provided in Figure~\ref{fig:appdeltaCPnew}, which 
shows excluded regions for three specific CP-violating phase values: $\delta_\text{CP}^{1(3)4} = 0$ (red), 
$\pi/2$ (green), and $\pi$ (blue). The dramatic variation in sensitivity with the CP-violating phase 
reveals the importance of these parameters in sterile neutrino searches. The strong dependence on the CP-violating phases can be understood by examining the
oscillation probability. To ease the discussion, we consider  the case of democratic active-sterile mixings, i.e. $\theta_{14}=\theta_{15} = \theta_{24} = \theta_{25}=\theta_{34} = \theta_{35} = \theta$ In this case, the appearance probabilities become:
\begin{equation}
\begin{split}
P_{\nu_{\mu}\rightarrow\nu_{e}} \simeq 4\theta^4 \Biggl[ 1 
+\frac{1}{2}\cos\delta_{14}  +\frac{1}{2}\cos\left(\Delta_{54}+\delta_{14}\right)  + {\cal O}(\theta^2)\Biggr].
\end{split}
\label{eq:prob_dis_app}
\end{equation}
\begin{equation}
\begin{split}
P_{\nu_{\mu}\rightarrow\nu_{\tau}} \simeq 4\theta^4 \Biggl[ 1 
+\frac{1}{2}\cos\delta_{34}  +\frac{1}{2}\cos\left(\Delta_{54}+\delta_{34}\right)  + {\cal O}(\theta^2)\Biggr].
\end{split}
\label{eq:prob_dis_app_tau}
\end{equation}

These expressions reveal the opposite behavior for $\delta_\text{CP}^{1(3)4}$ and 
$\delta_\text{CP}^{1(3)4} + \pi$, explaining the dramatic sensitivity variations 
observed in Figure~\ref{fig:appdeltaCPnew}.
In the small $\Delta_{54}$ limit, the probability depends critically on $\cos\delta_{1(3)4}$. 
When $\delta_{1(3)4}=0$, both cosine terms in Eq.~\eqref{eq:prob_dis_app} contribute 
constructively, maximizing the appearance probability and enabling the strongest 
experimental constraints (red contours). Conversely, when $\delta_{1(3)4}=\pi$, 
destructive interference leads to dramatic suppression of the appearance signal, 
with the probability approaching zero. This suppression renders the mixing parameters 
essentially unobservable, leading to the much weaker constraints shown by the blue 
contours in Figure~\ref{fig:appdeltaCPnew}. As $\Delta_{54}$ increases, the cancellation becomes less perfect due to the energy 
dependence of the oscillation phases, but the overall suppression effect persists.

%\clearpage

\section{Summary and conclusions}
\label{sec:conclusions}
The DUNE experiment is poised to significantly advance our understanding of neutrino oscillations and explore physics beyond the Standard Model, including the potential existence of sterile neutrinos motivated by persistent experimental anomalies. In this work, we have investigated DUNE's sensitivity to sterile neutrino oscillations within a 3+(pseudo-Dirac pair) framework, with a particular focus on its realization in the Linear Inverse Seesaw model. This scenario introduces two nearly degenerate keV-scale sterile neutrino states with mass-squared splitting ($\Delta m^2_{54} \sim \text{eV}^2$), which lead to distinct oscillation signatures detectable at DUNE's near and far detectors. 
Our study shows a distinctive phenomenological trait of the 3+(pseudo-Dirac pair) framework, setting it apart from simpler sterile neutrino models.  We characterized the oscillation behavior across three regimes defined by the sterile-sterile mass splitting $\Delta m^2_{54}$: low-mass, resonant, and high-mass regimes. We show that in both low- and high-mass regimes, the contribution from this scenario amounts to a constant deviation from unitarity in the oscillation probability and therefore those limits can be considered as a low-scale source of non-unitarity of the active block of the leptonic mixing matrix. Notably, a key feature is that even in the low-mass regime ($\Delta m^2_{54} \to 0$), this model predicts persistent, observable effects at the near detector, contrasting with standard 3+1 oscillation scenarios, where sensitivity is lost in this limit. This arises from the averaging of rapid oscillations driven by the larger, keV-scale splittings ($\Delta m^2_{41}, \Delta m^2_{51}$), which manifest as a constant, non-zero offset in the probabilities. While the resonant regime offers enhanced sensitivity due to the $\Delta m^2_{54}$-driven oscillations in disappearance channels, these resonant terms contribute at second order in the relevant mixing parameters (see Eq.~\eqref{eq:dis_prob}). Consequently, the expected improvement in sensitivity within this resonant regime for disappearance channels is less pronounced than in scenarios where such contributions are at leading order. In the appearance channels, the resonant terms contribute destructively and the best sensitivity is reached for the low mass-splitting regime.

Leveraging these features, we compute the future sensitivity of DUNE by employing the complementarity of both near and far detectors. The ND generally outperforms the FD, with the exception of $\nu_e$ disappearance, where the FD provides better sensitivity due to contributions from the interference with standard three-flavor oscillation terms in the probability, which are negligible in the ND. For disappearance channels, we expect an improvement over present bounds only in muon neutrino disappearance, and only in the resonant regime, since the resonant effects from which DUNE will benefit the most appear only at second order. In appearance channels, for $\nu_e$ appearance, the sensitivity on $|U_{e4}U_{e5}U_{\mu4}U_{\mu5}|$ can be as stringent as $4.6 \times 10^{-6}$ in the high-mass $\Delta m^2_{54}$ regime (and $3.4 \times 10^{-6}$ in the low-mass regime); for $\nu_\tau$ appearance, limits on $|U_{\tau4}U_{\tau5}U_{\mu4}U_{\mu5}|$ can reach $1.9 \times 10^{-5}$ and $2.5 \times 10^{-5}$ in the respective low- and high-mass regimes. These represent significant improvements over current experimental constraints, as detailed in Table~\ref{tab:constraints_summary}. Furthermore, our analysis of the LISS model parameter space, consistent with established neutrino oscillation data, pinpoints viable regions that DUNE can directly probe.

Finally, we  have also explored the impact of new CP-violating phases associated with the sterile sector, specifically $\delta_\text{CP}^{14}$ and $\delta_\text{CP}^{34}$ which enter in $\nu_e$ and $\nu_{\tau}$ appearance, respectively. Our results show that these phases can dramatically alter DUNE's sensitivity to appearance channels. Depending on their values, destructive interference can take place, suppressing the signal by orders of magnitude, lowering the expected sensitivity. 

In summary, our findings indicate that DUNE experiment has a very strong substantial discovery potential for the 3+(pseudo-Dirac pair) sterile neutrino scenario, with the capability to explore extensive regions of its parameter space within the LISS model and to significantly improve upon current experimental limits across multiple channels, especially in appearance channels. 

\paragraph{Acknowledgments} 
J. P. P. is supported by grant  PID2022-\allowbreak 126224NB-\allowbreak C21 and  "Unit of Excellence Maria de Maeztu 2020-2023'' award to the ICC-UB CEX2019-000918-M  funded by MCIN/AEI/\allowbreak 10.13039/\allowbreak 501100011033. 
also supported by the European Union's through the
Horizon 2020 research and innovation program (Marie
Sk{\l}odowska-Curie grant agreement 860881-HIDDeN).
This project has received funding from the European Union’s Horizon Europe research and innovation programme under the Marie Skłodowska-Curie Staff Exchange grant agreement No 101086085 – ASYMMETRY.

\appendix

\section{Parametrization of the leptonic mixing matrix}\label{App:Parametrization_leptonic_mixing}

In this appendix we present the detailed parametrization of the $5\times5$ leptonic mixing matrix $U$. The mixing matrix can be parametrized by a series of rotations as follows:
\begin{equation}
U = R_{45} R_{35} R_{25} R_{15} R_{34} R_{24} R_{14} R_{23} R_{13} R_{12} \; \mathrm{diag}\Big(1, e^{i\varphi_2}, e^{i\varphi_3}, e^{i\varphi_4}, e^{i\varphi_5}\Big),
\end{equation}
where each $R_{ij}$ denotes a rotation in the $ij$ plane, characterized by the rotation angle $\theta_{ij}$ and the corresponding Dirac phase $\delta_{ij}$ and $\varphi_i$ are the Majorana phases. As an example, in the $5$-dimensional space, the rotation $R_{14}$ acts non-trivially only on the first and fourth components and is explicitly given by
\begin{equation}
R_{14} =
\begin{pmatrix}
\cos\theta_{14} & 0 & 0 & \sin\theta_{14}e^{-i\delta_{14}} & 0 \\
0 & 1 & 0 & 0 & 0 \\
0 & 0 & 1 & 0 & 0 \\
-\sin\theta_{14}e^{i\delta_{14}} & 0 & 0 & \cos\theta_{14} & 0 \\
0 & 0 & 0 & 0 & 1 \\
\end{pmatrix}.
\end{equation}
Analogous expressions hold for the other rotation matrices $R_{ij}$.

Since there are six Dirac phases in the 5 dimensional mixing case, four of them can be absorbed by a redefinition of the charged-lepton fields. In our parametrization we choose to eliminate the phases by setting
\begin{equation}
\delta_{12} = \delta_{23} = \delta_{24} = \delta_{45} = 0.
\end{equation}

For completeness, we now provide the mapping between the mixing parameters and the active-heavy mixing elements:

\begin{align}
U_{e4} &= e^{-i(\delta_{14}-\varphi_4)}\, \cos\theta_{15}\,\sin\theta_{14},\\[1mm]
U_{e5} &= e^{-i(\delta_{15}-\varphi_5)}\, \sin\theta_{15},\\[1mm]
U_{\mu4} &= e^{i\varphi_4}\Big[\cos\theta_{14}\,\cos\theta_{25}\,\sin\theta_{24} - e^{-i(\delta_{14}-\delta_{15}+\delta_{25})}\,\sin\theta_{14}\,\sin\theta_{15}\,\sin\theta_{25}\Big],\\[1mm]
U_{\mu5} &= e^{-i(\delta_{25}-\varphi_5)}\, \cos\theta_{15}\,\sin\theta_{25},\\[1mm]
U_{\tau4} &= e^{-i(\delta_{14}+\delta_{34}+\delta_{35}-\varphi_4)}\Big[
e^{i(\delta_{14}+\delta_{35})}\cos\theta_{14}\,\cos\theta_{24}\,\cos\theta_{35}\,\sin\theta_{34}\\[1mm]
&\quad\quad - e^{i(\delta_{15}+\delta_{34})}\cos\theta_{25}\,\sin\theta_{14}\,\sin\theta_{15}\,\sin\theta_{35} \\[1mm]
&\quad\quad - e^{i(\delta_{14}+\delta_{25}+\delta_{34})}\cos\theta_{14}\,\sin\theta_{24}\,\sin\theta_{25}\,\sin\theta_{35}\Big],\\[1mm]
U_{\tau5} &= e^{-i(\delta_{35}-\varphi_5)}\, \cos\theta_{15}\,\cos\theta_{25}\,\sin\theta_{35}.
\end{align}

\section{Neutrino oscillation data}\label{App:neutrino_oscillation_data}
Any extension of the SM described by the LISS model must be consistent with observed neutrino oscillation data, including the solar and atmospheric mass-squared differences and mixing angles. The latest global fit to neutrino oscillation data, presented in Ref.~\cite{Esteban:2024eli}, provides the following updated values at the $3\sigma$ confidence level: 

\begin{eqnarray}
0.275\leq\sin^2\theta_{12}\leq0.345,\quad
0.435\leq\sin^2\theta_{23}\leq0.585,\quad
0.02030\leq\sin^2\theta_{13}\leq0.02388,\\
6.92\leq\frac{\Delta m_{21}^2}{10^{-5}~\mathrm{eV}^2}\leq8.05,\quad
2.451\leq\frac{\Delta m_{31}^2}{10^{-3}~\mathrm{eV}^2}\leq2.578,
\hspace{2cm}
\end{eqnarray}
for normal ordering and
\begin{eqnarray}
0.275\leq\sin^2\theta_{12}\leq0.345,\quad
0.440\leq\sin^2\theta_{23}\leq0.584,\quad
0.02060\leq\sin^2\theta_{13}\leq0.02409,\\
6.92\leq\frac{\Delta m_{21}^2}{10^{-5}~\mathrm{eV}^2}\leq8.05,\quad
-2.547\leq\frac{\Delta m_{32}^2}{10^{-3}~\mathrm{eV}^2}\leq-2.421,
\hspace{1.7cm}
\end{eqnarray}
for inverted ordering. In our parameter scans, a given point of the LISS parameter space will be considered valid if it reproduces neutrino oscillation parameters within these $3\sigma$ ranges.

\section{\cm{Oscillations constraints for 3+(pseudo-Dirac pair)}}\label{App:Constraints}

\cm{In this appendix we describe in detail how the current constraints quoted in 
Table~\ref{tab:constraints_summary} have been obtained.

For the low- and high-mass regimes, as already discussed in 
Section~\ref{subsec:ND_theory}, the limits can be interpreted in terms of 
low-scale non-unitarity. This is analogous to the high-mass limit 
($\Delta m^2_{41}\gg 1~\text{eV}^2$) of the $3+1$ scenario, from which the 
corresponding bounds can be read directly. 

In the resonance regime, the situation is more subtle. For the appearance 
combinations $|U_{e4}U_{e5}U_{\mu4}U_{\mu5}|$ and 
$|U_{\tau4}U_{\tau5}U_{\mu4}U_{\mu5}|$, destructive interference in 
Eq.~\eqref{eq:app_prob} prevents any improvement with respect to the high-mass 
bounds. We therefore adopt a conservative prescription and take these limits to 
be weaker than, or at most equal to, the corresponding high-mass constraints. For the disappearance parameters $4|U_{\alpha4}U_{\alpha5}|^2$, 
Eq.~\eqref{eq:dis_prob} shows that the energy-dependent contributions driven by 
$\Delta m_{54}^2$ arise only at order $O(U^4)$, while the constant terms from the 
averaged-out $\Delta m_{41}^2$ and $\Delta m_{51}^2$ oscillations appear already 
at order $O(U^2)$. As a consequence, an improvement in the resonance region can 
only be observed if the additional $U^2$ suppression is compensated. A necessary 
condition for this to occur is
\[
U_{\text{RES},3+1} < U^2_{\text{High mass},3+1}\,,
\]
where $U_{\text{RES},3+1}$ denotes the maximum sensitivity in the resonance 
regime and $U^2_{\text{High mass},3+1}$ the sensitivity in the high-mass regime 
of the $3+1$ scenario. With the present leading bounds this condition is not 
satisfied, and therefore the effective constraints in the resonance regime are 
taken to be the same as those in the low- and high-mass regimes.}

\bibliographystyle{JHEP} 
\bibliography{Refs}

\providecommand{\href}[2]{#2}\begingroup\raggedright\begin{thebibliography}{10}

\bibitem{Esteban:2024eli}
I.~Esteban, M.~C. Gonzalez-Garcia, M.~Maltoni, I.~Martinez-Soler, J.~P.
  Pinheiro, and T.~Schwetz, {\it {NuFit-6.0: updated global analysis of
  three-flavor neutrino oscillations}},  {\em JHEP} {\bf 12} (2024) 216,
  [\href{http://arxiv.org/abs/2410.05380}{{\tt arXiv:2410.05380}}].

\bibitem{LSND:1995lje}
{\bf LSND} Collaboration, C.~Athanassopoulos et~al., {\it {Candidate events in
  a search for anti-muon-neutrino ---\ensuremath{>} anti-electron-neutrino
  oscillations}},  {\em Phys. Rev. Lett.} {\bf 75} (1995) 2650--2653,
  [\href{http://arxiv.org/abs/nucl-ex/9504002}{{\tt nucl-ex/9504002}}].

\bibitem{MiniBooNE:2020pnu}
{\bf MiniBooNE} Collaboration, A.~A. Aguilar-Arevalo et~al., {\it {Updated
  MiniBooNE neutrino oscillation results with increased data and new background
  studies}},  {\em Phys. Rev. D} {\bf 103} (2021), no.~5 052002,
  [\href{http://arxiv.org/abs/2006.16883}{{\tt arXiv:2006.16883}}].

\bibitem{Kaether:2010ag}
F.~Kaether, W.~Hampel, G.~Heusser, J.~Kiko, and T.~Kirsten, {\it {Reanalysis of
  the GALLEX solar neutrino flux and source experiments}},  {\em Phys. Lett. B}
  {\bf 685} (2010) 47--54, [\href{http://arxiv.org/abs/1001.2731}{{\tt
  arXiv:1001.2731}}].

\bibitem{Abdurashitov:2005tb}
J.~N. Abdurashitov et~al., {\it {Measurement of the response of a Ga solar
  neutrino experiment to neutrinos from an Ar-37 source}},  {\em Phys. Rev. C}
  {\bf 73} (2006) 045805, [\href{http://arxiv.org/abs/nucl-ex/0512041}{{\tt
  nucl-ex/0512041}}].

\bibitem{GALLEX:1997lja}
{\bf GALLEX} Collaboration, W.~Hampel et~al., {\it {Final results of the Cr-51
  neutrino source experiments in GALLEX}},  {\em Phys. Lett. B} {\bf 420}
  (1998) 114--126.

\bibitem{SAGE:2009eeu}
{\bf SAGE} Collaboration, J.~N. Abdurashitov et~al., {\it {Measurement of the
  solar neutrino capture rate with gallium metal. III: Results for the
  2002--2007 data-taking period}},  {\em Phys. Rev. C} {\bf 80} (2009) 015807,
  [\href{http://arxiv.org/abs/0901.2200}{{\tt arXiv:0901.2200}}].

\bibitem{SAGE:1998fvr}
{\bf SAGE} Collaboration, J.~N. Abdurashitov et~al., {\it {Measurement of the
  response of the Russian-American gallium experiment to neutrinos from a Cr-51
  source}},  {\em Phys. Rev. C} {\bf 59} (1999) 2246--2263,
  [\href{http://arxiv.org/abs/hep-ph/9803418}{{\tt hep-ph/9803418}}].

\bibitem{Barinov:2021asz}
V.~V. Barinov et~al., {\it {Results from the Baksan Experiment on Sterile
  Transitions (BEST)}},  {\em Phys. Rev. Lett.} {\bf 128} (2022), no.~23
  232501, [\href{http://arxiv.org/abs/2109.11482}{{\tt arXiv:2109.11482}}].

\bibitem{Gonzalez-Garcia:2024hmf}
M.~C. Gonzalez-Garcia, M.~Maltoni, and J.~P. Pinheiro, {\it {Solar model
  independent constraints on the sterile neutrino interpretation of the Gallium
  Anomaly}},  {\em Phys. Lett. B} {\bf 862} (2025) 139297,
  [\href{http://arxiv.org/abs/2411.16840}{{\tt arXiv:2411.16840}}].

\bibitem{Minkowski:1977sc}
P.~Minkowski, {\it {$\mu \to e\gamma$ at a Rate of One Out of $10^{9}$ Muon
  Decays?}},  {\em Phys. Lett. B} {\bf 67} (1977) 421--428.

\bibitem{Yanagida:1979as}
T.~Yanagida, {\it {Horizontal gauge symmetry and masses of neutrinos}},  {\em
  Conf. Proc. C} {\bf 7902131} (1979) 95--99.

\bibitem{Mohapatra:1979ia}
R.~N. Mohapatra and G.~Senjanovic, {\it {Neutrino Mass and Spontaneous Parity
  Nonconservation}},  {\em Phys. Rev. Lett.} {\bf 44} (1980) 912.

\bibitem{GellMann:1980vs}
M.~Gell-Mann, P.~Ramond, and R.~Slansky, {\it {Complex Spinors and Unified
  Theories}},  {\em Conf. Proc. C} {\bf 790927} (1979) 315--321,
  [\href{http://arxiv.org/abs/1306.4669}{{\tt arXiv:1306.4669}}].

\bibitem{Akhmedov:1995ip}
E.~K. Akhmedov, M.~Lindner, E.~Schnapka, and J.~W.~F. Valle, {\it {Left-right
  symmetry breaking in NJL approach}},  {\em Phys. Lett. B} {\bf 368} (1996)
  270--280, [\href{http://arxiv.org/abs/hep-ph/9507275}{{\tt hep-ph/9507275}}].

\bibitem{Barr:2003nn}
S.~M. Barr, {\it {A Different seesaw formula for neutrino masses}},  {\em Phys.
  Rev. Lett.} {\bf 92} (2004) 101601,
  [\href{http://arxiv.org/abs/hep-ph/0309152}{{\tt hep-ph/0309152}}].

\bibitem{Malinsky:2005bi}
M.~Malinsky, J.~C. Romao, and J.~W.~F. Valle, {\it {Novel supersymmetric
  $SO(10)$ seesaw mechanism}},  {\em Phys. Rev. Lett.} {\bf 95} (2005) 161801,
  [\href{http://arxiv.org/abs/hep-ph/0506296}{{\tt hep-ph/0506296}}].

\bibitem{Mohapatra:1986bd}
R.~N. Mohapatra and J.~W.~F. Valle, {\it {Neutrino Mass and Baryon Number
  Nonconservation in Superstring Models}},  {\em Phys. Rev. D} {\bf 34} (1986)
  1642.

\bibitem{Gonzalez-Garcia:1988okv}
M.~C. Gonzalez-Garcia and J.~W.~F. Valle, {\it {Fast Decaying Neutrinos and
  Observable Flavor Violation in a New Class of Majoron Models}},  {\em Phys.
  Lett. B} {\bf 216} (1989) 360--366.

\bibitem{Gavela:2009cd}
M.~B. Gavela, T.~Hambye, D.~Hernandez, and P.~Hernandez, {\it {Minimal Flavour
  Seesaw Models}},  {\em JHEP} {\bf 09} (2009) 038,
  [\href{http://arxiv.org/abs/0906.1461}{{\tt arXiv:0906.1461}}].

\bibitem{BERNABEU1987303}
J.~Bernabéu, A.~Santamaria, J.~Vidal, A.~Mendez, and J.~Valle, {\it Lepton
  flavour non-conservation at high energies in a superstring inspired standard
  model},  {\em Physics Letters B} {\bf 187} (1987), no.~3 303--308.

\bibitem{Mohapatra:1986aw}
R.~N. Mohapatra, {\it {Mechanism for Understanding Small Neutrino Mass in
  Superstring Theories}},  {\em Phys. Rev. Lett.} {\bf 56} (1986) 561--563.

\bibitem{Abada:2015rta}
A.~Abada, G.~Arcadi, V.~Domcke, and M.~Lucente, {\it {Lepton number violation
  as a key to low-scale leptogenesis}},  {\em JCAP} {\bf 11} (2015) 041,
  [\href{http://arxiv.org/abs/1507.06215}{{\tt arXiv:1507.06215}}].

\bibitem{Wolfenstein:1981rk}
L.~Wolfenstein, {\it {CP Properties of Majorana Neutrinos and Double beta
  Decay}},  {\em Phys. Lett. B} {\bf 107} (1981) 77--79.

\bibitem{Petcov:1982ya}
S.~T. Petcov, {\it {On Pseudodirac Neutrinos, Neutrino Oscillations and
  Neutrinoless Double beta Decay}},  {\em Phys. Lett. B} {\bf 110} (1982)
  245--249.

\bibitem{Valle:1983dk}
J.~W.~F. Valle and M.~Singer, {\it {Lepton Number Violation With Quasi Dirac
  Neutrinos}},  {\em Phys. Rev. D} {\bf 28} (1983) 540.

\bibitem{Kobayashi:2000md}
M.~Kobayashi and C.~S. Lim, {\it {Pseudo Dirac scenario for neutrino
  oscillations}},  {\em Phys. Rev. D} {\bf 64} (2001) 013003,
  [\href{http://arxiv.org/abs/hep-ph/0012266}{{\tt hep-ph/0012266}}].

\bibitem{Abazajian:2012pn}
K.~N. Abazajian and M.~Kaplinghat, {\it {Detection of a Gamma-Ray Source in the
  Galactic Center Consistent with Extended Emission from Dark Matter
  Annihilation and Concentrated Astrophysical Emission}},  {\em Phys. Rev. D}
  {\bf 86} (2012) 083511, [\href{http://arxiv.org/abs/1207.6047}{{\tt
  arXiv:1207.6047}}]. [Erratum: Phys.Rev.D 87, 129902 (2013)].

\bibitem{Abazajian:2017tcc}
K.~N. Abazajian, {\it {Sterile neutrinos in cosmology}},  {\em Phys. Rept.}
  {\bf 711-712} (2017) 1--28, [\href{http://arxiv.org/abs/1705.01837}{{\tt
  arXiv:1705.01837}}].

\bibitem{Suliga:2019bsq}
A.~M. Suliga, I.~Tamborra, and M.-R. Wu, {\it {Tau lepton asymmetry by sterile
  neutrino emission -- Moving beyond one-zone supernova models}},  {\em JCAP}
  {\bf 12} (2019) 019, [\href{http://arxiv.org/abs/1908.11382}{{\tt
  arXiv:1908.11382}}].

\bibitem{Suliga:2020vpz}
A.~M. Suliga, I.~Tamborra, and M.-R. Wu, {\it {Lifting the core-collapse
  supernova bounds on keV-mass sterile neutrinos}},  {\em JCAP} {\bf 08} (2020)
  018, [\href{http://arxiv.org/abs/2004.11389}{{\tt arXiv:2004.11389}}].

\bibitem{Smirnov:2006bu}
A.~Y. Smirnov and R.~Zukanovich~Funchal, {\it {Sterile neutrinos: Direct mixing
  effects versus induced mass matrix of active neutrinos}},  {\em Phys. Rev. D}
  {\bf 74} (2006) 013001, [\href{http://arxiv.org/abs/hep-ph/0603009}{{\tt
  hep-ph/0603009}}].

\bibitem{Kusenko:2009up}
A.~Kusenko, {\it {Sterile neutrinos: The Dark side of the light fermions}},
  {\em Phys. Rept.} {\bf 481} (2009) 1--28,
  [\href{http://arxiv.org/abs/0906.2968}{{\tt arXiv:0906.2968}}].

\bibitem{Drewes:2016upu}
M.~Drewes et~al., {\it {A White Paper on keV Sterile Neutrino Dark Matter}},
  {\em JCAP} {\bf 01} (2017) 025, [\href{http://arxiv.org/abs/1602.04816}{{\tt
  arXiv:1602.04816}}].

\bibitem{Loewenstein:2008yi}
M.~Loewenstein, A.~Kusenko, and P.~L. Biermann, {\it {New Limits on Sterile
  Neutrinos from Suzaku Observations of the Ursa Minor Dwarf Spheroidal
  Galaxy}},  {\em Astrophys. J.} {\bf 700} (2009) 426--435,
  [\href{http://arxiv.org/abs/0812.2710}{{\tt arXiv:0812.2710}}].

\bibitem{Loewenstein:2012px}
M.~Loewenstein and A.~Kusenko, {\it {Dark Matter Search Using XMM-Newton
  Observations of Willman 1}},  {\em Astrophys. J.} {\bf 751} (2012) 82,
  [\href{http://arxiv.org/abs/1203.5229}{{\tt arXiv:1203.5229}}].

\bibitem{Kawasaki:2000en}
M.~Kawasaki, K.~Kohri, and N.~Sugiyama, {\it {MeV scale reheating temperature
  and thermalization of neutrino background}},  {\em Phys. Rev. D} {\bf 62}
  (2000) 023506, [\href{http://arxiv.org/abs/astro-ph/0002127}{{\tt
  astro-ph/0002127}}].

\bibitem{Hernandez:2014fha}
P.~Hernandez, M.~Kekic, and J.~Lopez-Pavon, {\it {$N_{eff}$ in low-scale seesaw
  models versus the lightest neutrino mass}},  {\em Phys. Rev. D} {\bf 90}
  (2014), no.~6 065033, [\href{http://arxiv.org/abs/1406.2961}{{\tt
  arXiv:1406.2961}}].

\bibitem{Gelmini:2004ah}
G.~Gelmini, S.~Palomares-Ruiz, and S.~Pascoli, {\it {Low reheating temperature
  and the visible sterile neutrino}},  {\em Phys. Rev. Lett.} {\bf 93} (2004)
  081302, [\href{http://arxiv.org/abs/astro-ph/0403323}{{\tt
  astro-ph/0403323}}].

\bibitem{Gelmini:2008fq}
G.~Gelmini, E.~Osoba, S.~Palomares-Ruiz, and S.~Pascoli, {\it {MeV sterile
  neutrinos in low reheating temperature cosmological scenarios}},  {\em JCAP}
  {\bf 10} (2008) 029, [\href{http://arxiv.org/abs/0803.2735}{{\tt
  arXiv:0803.2735}}].

\bibitem{Bezrukov:2017ike}
F.~Bezrukov, A.~Chudaykin, and D.~Gorbunov, {\it {Hiding an elephant: heavy
  sterile neutrino with large mixing angle does not contradict cosmology}},
  {\em JCAP} {\bf 06} (2017) 051, [\href{http://arxiv.org/abs/1705.02184}{{\tt
  arXiv:1705.02184}}].

\bibitem{Dasgupta:2013zpn}
B.~Dasgupta and J.~Kopp, {\it {Cosmologically Safe eV-Scale Sterile Neutrinos
  and Improved Dark Matter Structure}},  {\em Phys. Rev. Lett.} {\bf 112}
  (2014), no.~3 031803, [\href{http://arxiv.org/abs/1310.6337}{{\tt
  arXiv:1310.6337}}].

\bibitem{Barry:2014ika}
J.~Barry, J.~Heeck, and W.~Rodejohann, {\it {Sterile neutrinos and right-handed
  currents in KATRIN}},  {\em JHEP} {\bf 07} (2014) 081,
  [\href{http://arxiv.org/abs/1404.5955}{{\tt arXiv:1404.5955}}].

\bibitem{Petraki:2007gq}
K.~Petraki and A.~Kusenko, {\it {Dark-matter sterile neutrinos in models with a
  gauge singlet in the Higgs sector}},  {\em Phys. Rev. D} {\bf 77} (2008)
  065014, [\href{http://arxiv.org/abs/0711.4646}{{\tt arXiv:0711.4646}}].

\bibitem{Dodelson:1993je}
S.~Dodelson and L.~M. Widrow, {\it {Sterile-neutrinos as dark matter}},  {\em
  Phys. Rev. Lett.} {\bf 72} (1994) 17--20,
  [\href{http://arxiv.org/abs/hep-ph/9303287}{{\tt hep-ph/9303287}}].

\bibitem{Shrock:1980vy}
R.~E. Shrock, {\it {New Tests For, and Bounds On, Neutrino Masses and Lepton
  Mixing}},  {\em Phys. Lett. B} {\bf 96} (1980) 159--164.

\bibitem{Shrock:1981wq}
R.~E. Shrock, {\it {General Theory of Weak Processes Involving Neutrinos. 2.
  Pure Leptonic Decays}},  {\em Phys. Rev. D} {\bf 24} (1981) 1275.

\bibitem{Atre:2009rg}
A.~Atre, T.~Han, S.~Pascoli, and B.~Zhang, {\it {The Search for Heavy Majorana
  Neutrinos}},  {\em JHEP} {\bf 05} (2009) 030,
  [\href{http://arxiv.org/abs/0901.3589}{{\tt arXiv:0901.3589}}].

\bibitem{Abada:2013aba}
A.~Abada, A.~M. Teixeira, A.~Vicente, and C.~Weiland, {\it {Sterile neutrinos
  in leptonic and semileptonic decays}},  {\em JHEP} {\bf 02} (2014) 091,
  [\href{http://arxiv.org/abs/1311.2830}{{\tt arXiv:1311.2830}}].

\bibitem{Abada:2017jjx}
A.~Abada, V.~De~Romeri, M.~Lucente, A.~M. Teixeira, and T.~Toma, {\it
  {Effective Majorana mass matrix from tau and pseudoscalar meson lepton number
  violating decays}},  {\em JHEP} {\bf 02} (2018) 169,
  [\href{http://arxiv.org/abs/1712.03984}{{\tt arXiv:1712.03984}}].

\bibitem{Deppisch:2015qwa}
F.~F. Deppisch, P.~S. Bhupal~Dev, and A.~Pilaftsis, {\it {Neutrinos and
  Collider Physics}},  {\em New J. Phys.} {\bf 17} (2015), no.~7 075019,
  [\href{http://arxiv.org/abs/1502.06541}{{\tt arXiv:1502.06541}}].

\bibitem{Alekhin:2015byh}
S.~Alekhin et~al., {\it {A facility to Search for Hidden Particles at the CERN
  SPS: the SHiP physics case}},  {\em Rept. Prog. Phys.} {\bf 79} (2016),
  no.~12 124201, [\href{http://arxiv.org/abs/1504.04855}{{\tt
  arXiv:1504.04855}}].

\bibitem{Antel:2023hkf}
C.~Antel et~al., {\it {Feebly-interacting particles: FIPs 2022 Workshop
  Report}},  {\em Eur. Phys. J. C} {\bf 83} (2023), no.~12 1122,
  [\href{http://arxiv.org/abs/2305.01715}{{\tt arXiv:2305.01715}}].

\bibitem{KATRIN:2001ttj}
{\bf KATRIN} Collaboration, A.~Osipowicz et~al., {\it {KATRIN: A Next
  generation tritium beta decay experiment with sub-eV sensitivity for the
  electron neutrino mass. Letter of intent}},
  \href{http://arxiv.org/abs/hep-ex/0109033}{{\tt hep-ex/0109033}}.

\bibitem{Mertens:2014nha}
S.~Mertens, T.~Lasserre, S.~Groh, G.~Drexlin, F.~Glueck, A.~Huber, A.~W.~P.
  Poon, M.~Steidl, N.~Steinbrink, and C.~Weinheimer, {\it {Sensitivity of
  Next-Generation Tritium Beta-Decay Experiments for keV-Scale Sterile
  Neutrinos}},  {\em JCAP} {\bf 02} (2015) 020,
  [\href{http://arxiv.org/abs/1409.0920}{{\tt arXiv:1409.0920}}].

\bibitem{Boyarsky:2018tvu}
A.~Boyarsky, M.~Drewes, T.~Lasserre, S.~Mertens, and O.~Ruchayskiy, {\it
  {Sterile neutrino Dark Matter}},  {\em Prog. Part. Nucl. Phys.} {\bf 104}
  (2019) 1--45, [\href{http://arxiv.org/abs/1807.07938}{{\tt
  arXiv:1807.07938}}].

\bibitem{Abada:2018qok}
A.~Abada, A.~Hern\'andez-Cabezudo, and X.~Marcano, {\it {Beta and Neutrinoless
  Double Beta Decays with KeV Sterile Fermions}},  {\em JHEP} {\bf 01} (2019)
  041, [\href{http://arxiv.org/abs/1807.01331}{{\tt arXiv:1807.01331}}].

\bibitem{DUNE:2020ypp}
{\bf DUNE} Collaboration, B.~Abi et~al., {\it {Deep Underground Neutrino
  Experiment (DUNE), Far Detector Technical Design Report, Volume II: DUNE
  Physics}},  \href{http://arxiv.org/abs/2002.03005}{{\tt arXiv:2002.03005}}.

\bibitem{DUNE:2020jqi}
{\bf DUNE} Collaboration, B.~Abi et~al., {\it {Long-baseline neutrino
  oscillation physics potential of the DUNE experiment}},  {\em Eur. Phys. J.
  C} {\bf 80} (2020), no.~10 978, [\href{http://arxiv.org/abs/2006.16043}{{\tt
  arXiv:2006.16043}}].

\bibitem{Abe:2015zbg}
{\bf Hyper-Kamiokande Proto-} Collaboration, K.~Abe et~al., {\it {Physics
  potential of a long-baseline neutrino oscillation experiment using a J-PARC
  neutrino beam and Hyper-Kamiokande}},  {\em PTEP} {\bf 2015} (2015) 053C02,
  [\href{http://arxiv.org/abs/1502.05199}{{\tt arXiv:1502.05199}}].

\bibitem{Abe:2018uyc}
{\bf Hyper-Kamiokande} Collaboration, K.~Abe et~al., {\it {Hyper-Kamiokande
  Design Report}},  \href{http://arxiv.org/abs/1805.04163}{{\tt
  arXiv:1805.04163}}.

\bibitem{nuPRISM:2014mzw}
{\bf nuPRISM} Collaboration, S.~Bhadra et~al., {\it {Letter of Intent to
  Construct a nuPRISM Detector in the J-PARC Neutrino Beamline}},
  \href{http://arxiv.org/abs/1412.3086}{{\tt arXiv:1412.3086}}.

\bibitem{Hyper-Kamiokande:2018ofw}
{\bf Hyper-Kamiokande} Collaboration, K.~Abe et~al., {\it {Hyper-Kamiokande
  Design Report}},  \href{http://arxiv.org/abs/1805.04163}{{\tt
  arXiv:1805.04163}}.

\bibitem{HyperKamiokande2025ESPP}
F.~D. Lodovico, {\it The hyper-kamiokande experiment: input to the update of
  the european strategy for particle physics},  in {\em Input to the European
  Strategy for Particle Physics - 2026 update}, 2025.

\bibitem{T2K:2011qtm}
{\bf T2K} Collaboration, K.~Abe et~al., {\it {The T2K Experiment}},  {\em Nucl.
  Instrum. Meth. A} {\bf 659} (2011) 106--135,
  [\href{http://arxiv.org/abs/1106.1238}{{\tt arXiv:1106.1238}}].

\bibitem{DUNE:2020lwj}
{\bf DUNE} Collaboration, B.~Abi et~al., {\it {Deep Underground Neutrino
  Experiment (DUNE), Far Detector Technical Design Report, Volume I
  Introduction to DUNE}},  {\em JINST} {\bf 15} (2020), no.~08 T08008,
  [\href{http://arxiv.org/abs/2002.02967}{{\tt arXiv:2002.02967}}].

\bibitem{Huber:2004ka}
P.~Huber, M.~Lindner, and W.~Winter, {\it {Simulation of long-baseline neutrino
  oscillation experiments with GLoBES (General Long Baseline Experiment
  Simulator)}},  {\em Comput. Phys. Commun.} {\bf 167} (2005) 195,
  [\href{http://arxiv.org/abs/hep-ph/0407333}{{\tt hep-ph/0407333}}].

\bibitem{Huber:2007ji}
P.~Huber, J.~Kopp, M.~Lindner, M.~Rolinec, and W.~Winter, {\it {New features in
  the simulation of neutrino oscillation experiments with GLoBES 3.0: General
  Long Baseline Experiment Simulator}},  {\em Comput. Phys. Commun.} {\bf 177}
  (2007) 432--438, [\href{http://arxiv.org/abs/hep-ph/0701187}{{\tt
  hep-ph/0701187}}].

\bibitem{Coloma:2021uhq}
P.~Coloma, J.~L\'opez-Pav\'on, S.~Rosauro-Alcaraz, and S.~Urrea, {\it {New
  physics from oscillations at the DUNE near detector, and the role of
  systematic uncertainties}},  {\em JHEP} {\bf 08} (2021) 065,
  [\href{http://arxiv.org/abs/2105.11466}{{\tt arXiv:2105.11466}}].

\bibitem{Blennow:2025qgd}
M.~Blennow, P.~Coloma, E.~Fern\'andez-Mart\'\i{}nez,
  J.~Hern\'andez-Garc\'\i{}a, J.~L\'opez-Pav\'on, X.~Marcano, D.~Naredo-Tuero,
  and S.~Urrea, {\it {Misconceptions in neutrino oscillations in presence of
  non-unitary mixing}},  {\em Nucl. Phys. B} {\bf 1017} (2025) 116944,
  [\href{http://arxiv.org/abs/2502.19480}{{\tt arXiv:2502.19480}}].

\bibitem{Blennow:2023mqx}
M.~Blennow, E.~Fern\'andez-Mart\'\i{}nez, J.~Hern\'andez-Garc\'\i{}a,
  J.~L\'opez-Pav\'on, X.~Marcano, and D.~Naredo-Tuero, {\it {Bounds on lepton
  non-unitarity and heavy neutrino mixing}},  {\em JHEP} {\bf 08} (2023) 030,
  [\href{http://arxiv.org/abs/2306.01040}{{\tt arXiv:2306.01040}}].

\bibitem{tHooft:1979rat}
G.~'t~Hooft, {\it {Naturalness, chiral symmetry, and spontaneous chiral
  symmetry breaking}},  {\em NATO Sci. Ser. B} {\bf 59} (1980) 135--157.

\bibitem{Dev:2012sg}
P.~S.~B. Dev and A.~Pilaftsis, {\it {Minimal Radiative Neutrino Mass Mechanism
  for Inverse Seesaw Models}},  {\em Phys. Rev. D} {\bf 86} (2012) 113001,
  [\href{http://arxiv.org/abs/1209.4051}{{\tt arXiv:1209.4051}}].

\bibitem{Lopez-Pavon:2012yda}
J.~Lopez-Pavon, S.~Pascoli, and C.-f. Wong, {\it {Can heavy neutrinos dominate
  neutrinoless double beta decay?}},  {\em Phys. Rev. D} {\bf 87} (2013), no.~9
  093007, [\href{http://arxiv.org/abs/1209.5342}{{\tt arXiv:1209.5342}}].

\bibitem{Casas:2001sr}
J.~A. Casas and A.~Ibarra, {\it {Oscillating neutrinos and $\mu \to e
  \gamma$}},  {\em Nucl. Phys. B} {\bf 618} (2001) 171--204,
  [\href{http://arxiv.org/abs/hep-ph/0103065}{{\tt hep-ph/0103065}}].

\bibitem{Abada:2018sfh}
A.~Abada, N.~Bernal, M.~Losada, and X.~Marcano, {\it {Inclusive Displaced
  Vertex Searches for Heavy Neutral Leptons at the LHC}},  {\em JHEP} {\bf 01}
  (2019) 093, [\href{http://arxiv.org/abs/1807.10024}{{\tt arXiv:1807.10024}}].

\bibitem{Abada:2022wvh}
A.~Abada, P.~Escribano, X.~Marcano, and G.~Piazza, {\it {Collider searches for
  heavy neutral leptons: beyond simplified scenarios}},  {\em Eur. Phys. J. C}
  {\bf 82} (2022), no.~11 1030, [\href{http://arxiv.org/abs/2208.13882}{{\tt
  arXiv:2208.13882}}].

\bibitem{Drewes:2022akb}
M.~Drewes, J.~Klari\'c, and J.~L\'opez-Pav\'on, {\it {New benchmark models for
  heavy neutral lepton searches}},  {\em Eur. Phys. J. C} {\bf 82} (2022),
  no.~12 1176, [\href{http://arxiv.org/abs/2207.02742}{{\tt
  arXiv:2207.02742}}].

\bibitem{DUNE:2021tad}
{\bf DUNE} Collaboration, V.~Hewes et~al., {\it {Deep Underground Neutrino
  Experiment (DUNE) Near Detector Conceptual Design Report}},  {\em
  Instruments} {\bf 5} (2021), no.~4 31,
  [\href{http://arxiv.org/abs/2103.13910}{{\tt arXiv:2103.13910}}].

\bibitem{DUNE:2021cuw}
{\bf DUNE} Collaboration, B.~Abi et~al., {\it {Experiment Simulation
  Configurations Approximating DUNE TDR}},
  \href{http://arxiv.org/abs/2103.04797}{{\tt arXiv:2103.04797}}.

\bibitem{Coyle:2025xjk}
N.~M. Coyle, S.~W. Li, and P.~A.~N. Machado, {\it {Neutrino-Nucleus Cross
  Section Impacts on Neutrino Oscillation Measurements}},
  \href{http://arxiv.org/abs/2502.19467}{{\tt arXiv:2502.19467}}.

\bibitem{Miranda:2018yym}
O.~G. Miranda, P.~Pasquini, M.~T\'ortola, and J.~W.~F. Valle, {\it {Exploring
  the Potential of Short-Baseline Physics at Fermilab}},  {\em Phys. Rev. D}
  {\bf 97} (2018), no.~9 095026, [\href{http://arxiv.org/abs/1802.02133}{{\tt
  arXiv:1802.02133}}].

\bibitem{Meloni:2018xnk}
D.~Meloni, {\it {On the systematic uncertainties in DUNE and their role in New
  Physics studies}},  {\em JHEP} {\bf 08} (2018) 028,
  [\href{http://arxiv.org/abs/1805.01747}{{\tt arXiv:1805.01747}}].

\bibitem{Alion:2016uaj}
{\bf DUNE} Collaboration, T.~Alion et~al., {\it {Experiment Simulation
  Configurations Used in DUNE CDR}},
  \href{http://arxiv.org/abs/1606.09550}{{\tt arXiv:1606.09550}}.

\bibitem{Andreopoulos:2015wxa}
C.~Andreopoulos, C.~Barry, S.~Dytman, H.~Gallagher, T.~Golan, R.~Hatcher,
  G.~Perdue, and J.~Yarba, {\it {The GENIE Neutrino Monte Carlo Generator:
  Physics and User Manual}},  \href{http://arxiv.org/abs/1510.05494}{{\tt
  arXiv:1510.05494}}.

\bibitem{Tena-Vidal:2021rpu}
{\bf GENIE} Collaboration, J.~Tena-Vidal et~al., {\it {Neutrino-Nucleon
  Cross-Section Model Tuning in GENIE v3}},
  \href{http://arxiv.org/abs/2104.09179}{{\tt arXiv:2104.09179}}.

\bibitem{DeGouvea:2019kea}
A.~De~Gouv\^ea, K.~J. Kelly, G.~V. Stenico, and P.~Pasquini, {\it {Physics with
  Beam Tau-Neutrino Appearance at DUNE}},  {\em Phys. Rev. D} {\bf 100} (2019),
  no.~1 016004, [\href{http://arxiv.org/abs/1904.07265}{{\tt
  arXiv:1904.07265}}].

\bibitem{Zyla:2020zbs}
{\bf Particle Data Group} Collaboration, P.~Zyla et~al., {\it {Review of
  Particle Physics}},  {\em PTEP} {\bf 2020} (2020), no.~8 083C01.

\bibitem{Wilks:1938dza}
S.~S. Wilks, {\it {The Large-Sample Distribution of the Likelihood Ratio for
  Testing Composite Hypotheses}},  {\em Annals Math. Statist.} {\bf 9} (1938),
  no.~1 60--62.

\bibitem{Goldhagen:2021kxe}
K.~Goldhagen, M.~Maltoni, S.~E. Reichard, and T.~Schwetz, {\it {Testing sterile
  neutrino mixing with present and future solar neutrino data}},  {\em Eur.
  Phys. J. C} {\bf 82} (2022), no.~2 116,
  [\href{http://arxiv.org/abs/2109.14898}{{\tt arXiv:2109.14898}}].

\bibitem{MINOS:2020iqj}
{\bf MINOS+, Daya Bay} Collaboration, P.~Adamson et~al., {\it {Improved
  Constraints on Sterile Neutrino Mixing from Disappearance Searches in the
  MINOS, MINOS+, Daya Bay, and Bugey-3 Experiments}},  {\em Phys. Rev. Lett.}
  {\bf 125} (2020), no.~7 071801, [\href{http://arxiv.org/abs/2002.00301}{{\tt
  arXiv:2002.00301}}].

\bibitem{NOMAD:2001xxt}
{\bf NOMAD} Collaboration, P.~Astier et~al., {\it {Final NOMAD results on
  muon-neutrino ---\ensuremath{>} tau-neutrino and electron-neutrino
  ---\ensuremath{>} tau-neutrino oscillations including a new search for
  tau-neutrino appearance using hadronic tau decays}},  {\em Nucl. Phys. B}
  {\bf 611} (2001) 3--39, [\href{http://arxiv.org/abs/hep-ex/0106102}{{\tt
  hep-ex/0106102}}].

\end{thebibliography}\endgroup
\end{document}